\documentclass[draft]{agujournal2019}
\usepackage{amsmath}
\usepackage{amssymb}
\usepackage{url} 
\usepackage{lineno}
\usepackage[inline]{trackchanges} 
\usepackage{soul}
\draftfalse

\journalname{JGR: Solid Earth}

\begin{document}

%
%


\title{Depth and thickness of tectonic tremor in the northeastern Olympic Peninsula}

%
%




\authors{A. Ducellier\affil{1}, K. C. Creager\affil{1}}


\affiliation{1}{University of Washington}




\correspondingauthor{Ariane Ducellier}{ducela@uw.edu}




\begin{keypoints}
\item We use seismic data from small-aperture arrays in the Olympic Peninsula, and tremor epicenters determined by a previous study. 
\item We compute the depth of the tremor from S minus P times determined from stacking horizontal-to-vertical cross-correlations of seismic data.
\item Tremor is located close to the plate boundary in a region no more than 2-3 kilometers thick.
\end{keypoints}

%
%

%
%


\begin{abstract}
Tectonic tremor has been explained as a swarm of low-frequency earthquakes, which are located on a narrow fault at the plate boundary. However, due to the lack of clear impulsive phases in the tremor signal, it is difficult to determine the depth of the tremor source with great precision. The thickness of the tremor region is also not well constrained. The tremor may be located on a narrow fault as the low-frequency earthquakes appear to be, or distributed over a few kilometers wide low shear-wave velocity layer in the upper oceanic crust, which is thought to be a region with high pore-fluid pressure. Lag times of peaks in the cross-correlation of the horizontal and vertical components of tremor seismograms, recorded by small-aperture arrays in the Olympic Peninsula, Washington, are interpreted to be S minus P times.  Tremor depths are estimated from these S minus P times using epicenters from a previous study using a multibeam backprojection method. The tremor is located close to the plate boundary in a region no more than 2-3 kilometers thick and is very close to the depths of low-frequency earthquakes. The tremor is distributed over a wider depth range than the low-frequency earthquakes. However, due to the uncertainty on the depth, it is difficult to conclude whether the source of the tremor is located at the top of the subducting oceanic crust, in the lower continental crust just above the plate boundary, or in a narrow zone at the plate boundary.
\end{abstract}

\section*{Plain Language Summary}
Tectonic tremor is a long, low amplitude seismic signal, with emergent onsets, and an absence of clear impulsive phases. It has been explained as a swarm of low-frequency earthquakes, which are located on a narrow fault at the plate boundary. It is therefore assumed that the source of the tectonic tremor is located close to the plate boundary. However, due to the lack of clear impulsive phases in the tremor signal, it is difficult to determine the depth of the tremor source and the distance of the source to the plate interface with great precision. The thickness of the tremor region is not well constrained either. The tremor may be located on a narrow fault like the low-frequency earthquakes, or distributed through a few kilometers wide low shear-wave velocity layer in the upper oceanic crust, which is thought to be a region with high pore-fluid pressure. In this paper, we show that the tremor is located close to the plate boundary in a region no more than 2-3 kilometers thick and is very close to the depths of low-frequency earthquakes. Knowing the depth of the tremor source will help understand the mechanism that produces the tremor.

%
%

%


%
%
%
%

\section{Introduction}

Tremor is a long (several seconds to many minutes), low amplitude seismic signal, with emergent onsets, and an absence of clear impulsive phases. Tectonic tremor has been explained as a swarm of small, low-frequency earthquakes (LFEs) ~\cite{SHE_2007_nature}, that is small magnitude earthquakes (M $\sim$ 1) where frequency content (1-10 Hz) is lower than for ordinary small earthquakes (up to 20 Hz). The source of the LFEs is located on or near the plate boundary, and their focal mechanisms represent shear slip on a low-angle thrust fault dipping in the same direction as the plate interface ~\cite{IDE_2007_GRL, ROY_2014}. The polarization of tremor in northern Cascadia is also consistent with slip on the pate boundary in the direction of relative plate motion ~\cite{WEC_2007}.  Due to the lack of clear impulsive phases in the tremor signal, it is difficult to determine the depth of the tremor source with the same precision. In subduction zones such as Nankai and Cascadia, tectonic tremor observations are spatially and temporally correlated with geodetically observed slow slip  ~\cite{OBA_2002, ROG_2003}. Due to this correlation, these paired phenomena have been called Episodic Tremor and Slip (ETS). \\

The occurrence of tremor seems to be linked to low effective normal stress or high fluid pressure near the location of the tremor. Indeed, ~\citeA{SHE_2006} have observed a high P-wave to S-wave velocity ratio in the subducting oceanic crust near the location of the LFEs in western Shikoku, Japan. In Cascadia, ~\citeA{AUD_2009} have computed teleseismic receiver functions in Vancouver Island, and analyzed the delay times between the forward-scattered P-to-S, and back-scattered P-to-S and S-to-S conversions at two seismic reflectors identified as the top and bottom of the oceanic crust. It allowed them to compute the P-to-S velocity ratio of the oceanic crust layer and the S-wave velocity contrast at both interfaces. The very low Poisson's ratio of the layer could not be explained by the mineral composition, and they interpreted it as evidence for high pore-fluid pressure. The link between tremor and high pore fluid pressure is also supported by the influence of tidal cycles on the variations of tremor occurrence. \citeA{NAK_2008} noticed that tremor swarms often exhibit occurrences with a periodicity of about 12 or 24 h, and concluded that they are probably related to Earth tides. Their occurrence is also well correlated with time evolution of Coulomb failure stress (CFS) and CFS rate. However, tremor occurrences are advanced by a few hours relative to CFS, from which they conclude that a simple Coulomb threshold model is not sufficient to explain tremor occurrence. This discrepancy can be explained by using a rate- and state-dependent friction law. ~\citeA{THO_2009} have also observed that tremor occurrence on the deep San Andreas fault are correlated with small, tidal shear stress changes. They explain it by a very weak fault zone with low effective normal stress, probably due to near-lithostatic pore pressures at the depth of the tremor source region. \\

Two scenarii could explain the generation of tectonic tremor by highly pressured fluids ~\cite{SHE_2006}. A first possibility is that tremor is generated by the movement of fluids at depth, either by hydraulic fracturing or by coupling between the rock and fluid flow. The accompanying slip could be triggered by the same fluid movement that generates the tremor or, alternatively, the fluid flow could be a response to changes in stress and strain induced by the accompanying slip. The second possibility is that tremor is generated by slow otherwise aseismic shear slip on the plate interface as slip locally accelerates owing to the effects of geometric or physical irregularities on the plate interface. Fluids would then play an auxiliary role, altering the conditions on the plate interface to enable transient slip events, without generating seismic waves directly. \\

The source of the fluids could be the dehydration of hydrous minerals within the subducting oceanic crust ~\cite{SHE_2006}. ~\citeA{FAG_2011} computed the equilibrium mineral assemblages at different P-T conditions, and compared it to the P-T path of the subducting oceanic crust in Shikoku and Cascadia. They noted that for most of the P-T path, there are no dehydration reactions and the slab remains fluid-absent, except for depths between 30 and 35 km for Shikoku, and depths between 30 and 40 km for Cascadia, where the mineral model predicts significant water release. These depth ranges coincide with the depth range where tremor has been observed. They concluded that abundant tremor activity requires metamorphic conditions where localized dehydration occurs during subduction, which explains why the generation of tectonic tremor is restricted to a small range of depth along the plate boundary. In subduction zones where dehydration reactions are more widely distributed, there would be a more diffuse pattern of tremor activity that would be harder to detect. \\

Large amounts of fluids could be available at the fore-arc mantle corner. First, the bending of the subducting plate at the ocean trench may introduce water in the upper oceanic mantle, resulting in extensive serpentinization ~\cite{HYN_2015}. Second, the serpentinization of the fore-arc mantle corner may decrease the vertical permeability of the boundary between the oceanic plate and the overriding continental crust while keeping a high permeability parallel to the fault. It would thus channel all the fluid updip in the subducting oceanic crust, and explain the sharp velocity contrast observed by ~\citeA{AUD_2009} on top of the oceanic crust layer. At greater depth, the large volume reduction and water release accompanying eclogitization in the subducted oceanic crust, and the large volume expansion accompanying serpentinization in the mantle wedge, could increase the permeability of the plate boundary through fracture generation. A possible cause of ETS events could thus be periodic cycles of steady pore-fluid pressure build-up from dehydration of subducted oceanic crust, fluid release from fracturing of the interface during ETS, and subsequent precipitation sealing of the plate boundary ~\cite{AUD_2009}. \\

Whereas the position of the subduction zone ETS does not seem to coincide with a specific temperature or dehydration reaction ~\cite{PEA_2009}, there seems to be a good coincidence between the location of the fore-arc mantle corner, and the location of ETS ~\cite{HYN_2015}. The generation of slow slip and tectonic tremor could then be related to the presence of quartz in the overriding continental crust. Using receiver functions of teleseismic body waves, and data from the literature, ~\citeA{AUD_2014} observed that the recurrence time of slow earthquakes increases linearly with the Vp / Vs ratio of the forearc crust. They also noticed that along a margin-perpendicular profile from northern Cascadia, the Vp / Vs ratio of the forearc, and the recurrence time of ETS events, decrease with increasing depth. Likewise, ~\citeA{HYN_2015}  pointed out that the deep fore-arc crust has a very low Poisson's ratio (less than 0.22), and that the only mineral with a very low Poisson's ratio is quartz (about 0.1), which led them to conclude that there may be a significant amount of quartz (about 10\% by volume) in the deep fore-arc crust above the fore-arc mantle. ~\citeA{AUD_2014} explained the presence of quartz in the forearc by the enrichment of forearc minerals in fluid-dissolved silica derived from the dehydration of the down-going slab. As the solubility of silica increases with temperature, fluids generated at depth and rising up the subduction channel should be rich in silica ~\cite{HYN_2015}. \\

Quartz veins have indeed been observed in exhumed subduction zones. For instance, ~\citeA{FAG_2014} have studied an exhumed shear zone representing the subduction megathrust before its incorporation into the accretionary prism. They focused their study on a 30 m high by 80 m long cliff exposure where foliation has developed as a result of shearing along the subduction thrust interface. They identified two groups of quartz veins, foliation-parallel veins, and discordant veins, that must have formed for an extended time before, during, and after foliation development. They interpret the foliation-parallel veins as having been formed by viscous shear flow, and note that the shear stain rate due to the flow may be high enough to accommodate a slow slip strain rate of $~10^{-9} s^{-1}$, for a typical subduction thrust thickness of 30 m ~\cite{ROW_2013}. They interpret the discordant veins as having been formed by brittle deformation caused by locally elevated fluid pressure. The size of the structures where brittle deformation is observed (meters to hundreds of meters) is compatible with the size of the asperity rupturing during an LFE. Tremor and slow slip may thus be a manifestation of brittle-viscous deformation in the shear zone. \\

However, several constraints remain to be explained. ~\citeA{AUD_2014} estimated that the fluid flux required for the formation of quartz veins was two orders of magnitude greater than the fluid production rates estimated from the dehydration of the slab. They hypothesized that silica-saturated fluids may originate from the complete serpentinization of the mantle near the wedge corner. They suggested that higher temperature and quartz content at depth may lead to faster dissolution - precipitation processes and more frequent slip events. Their model could also explain the global variation in recurrence time, with mafic silica-poor regions having longer ETS recurrence times than felsic silica-rich regions. \\

Moreover, the zone with high low Poisson's ratio observed by ~\citeA{HYN_2015} has a large vertical extent (about 10 kilometers). If the whole zone is associated with quartz deposition and tectonic tremor generation, we should also observe a large vertical distribution of the source of the tremor. ~\citeA{KAO_2006} have used a Source Scanning Algorithm to detect and locate tremor, and have indeed located tremor in the continental crust, with a wide depth range of over 40 km. They noted that this wide depth range could not arise from either analysis uncertainties or a systematic bias in the velocity model they used. Uncertainties on the location of the tremor have been estimated by  ~\citeA{IDE_2012} at about 1.5 km in epicenter and 4.5 km in depth. A follow-up study by ~\citeA{KAO_2009} gave a thickness of the tremor zone of 5-10 km. This depth range is inconsistent with the depth of the LFEs, which have been located on a thin band at or near the plate interface with a rupture mechanism that corresponds to the thrust dip angle ~\cite{IDE_2007_GRL}. In the Olympic Peninsula, LFE families have been identified and located by ~\citeA{CHE_2017_JGR,CHE_2017_G3}, and all LFE families were located near the plate interface. Further study is thus needed to narrow the uncertainty on the depth of the source of the tremor, and verify whether tremor occurs in a wider zone than LFEs. \\

Several methods have been developed to detect and locate tectonic tremor or LFEs using the cross-correlation of seismic signals. The main idea is to find similar waveforms in two different seismic signals, which could correspond to a single tremor or LFE recorded at two different stations, or two different tremors or LFEs with the same source location but occurring at two different times and recorded by the same station. A first method consists in comparing the envelopes of seismograms at different stations ~\cite{OBA_2002, WEC_2008}, or directly the seismograms at different stations ~\cite{RUB_2013}. For instance, ~\citeA{WEC_2008} computed the cross-correlations of envelope seismograms for a set of 20 stations in western Washington and southern Vancouver Island. Then, they performed a grid search over all possible source locations to determine which one minimizes the difference between the maximum cross-correlation and the value of the correlogram at the lag time corresponding to the S-wave travel time difference between two stations. \\

A second method is based on the assumption that repeating tremor or LFEs with sources located nearby in space will have similar waveforms ~\cite{BOS_2012, ROY_2014, SHE_2006, SHE_2007_nature}. For instance, ~\citeA{BOS_2012} looked for LFEs by computing autocorrelations of 6-second long windows for each component of 7 stations in Vancouver Island. They then classified their LFE detections into 140 families. By stacking all waveforms of a given family, they obtained an LFE template for each family. They extended their templates by adding more stations and computing cross-correlations between station data and template waveforms. They used P- and S-travel-time picks to obtain a hypocenter for each LFE template. By observing the polarizations of the P- and S-waveforms of the LFE templates, they computed focal mechanisms and obtained a mixture of strike slip and thrust mechanisms, corresponding to a compressive stress field consistent with thrust faulting parallel to the plate interface. Further study showed that the average double couple solution is generally consistent with shallow thrusting in the direction of plate motion \cite{ROY_2014}. \\

Finally, a third method uses seismograms recorded across small-aperture arrays ~\cite{GHO_2010_GRL, LAR_2009}. For instance, ~\citeA{LAR_2009} stacked seismograms over all stations of the array for each component, and for three arrays in Cascadia. They then computed the cross-correlation between the horizontal and the vertical component, and found a distinct and persistent peak at a positive lag time, corresponding to the time between P-wave arrival on the vertical channel and S-wave arrival on the horizontal channels. Using a standard layered Earth model, and horizontal slowness estimated from array analysis, they computed the depths of the tremor sources. They located the sources near or at the plate interface, with a much better depth resolution than previous methods based on seismic signal envelopes, source scanning algorithm, or small-aperture arrays. They concluded that at least some of the tremor consisted in the repetition of LFEs as was the case in Shikoku. A drawback of the method was that it could be applied only to tremor located beneath an array, and coming from only one place for an extended period of time. \\

In this study, we extend on the method used by ~\citeA{LAR_2009} using the cross-correlation between horizontal and vertical components of seismic recordings to estimate the depth of the source of the tectonic tremor, and the depth extent of the region from which tremor originates. If indeed tremor is made of swarms of LFEs, and both represent the regions of deformation during ETS events, we would expect the thickness of the tremor to be the same as the thickness of the LFEs. If not, tremor may be occurring where LFEs are harder to detect because they are either smaller in amplitude, or spread out over continuous space and not as clearly repeating.

\section{Data}

The data were collected during the 2009-2011 Array of Arrays experiment. Eight small-aperture arrays were installed in the northeastern part of the Olympic Peninsula, Washington. The aperture of the arrays was about 1 km, and station spacing was a few hundred meters. Arrays typically had ten 3-component seismometers, augmented by an additional 10 vertical-only sensors during the 2010 ETS event.  The arrays were around 5 to 10 km apart from each other (Figure 1). Most of the arrays recorded data  for most of  a year, between June 2009 to September 2010, and captured the main August 2010 ETS event. The arrays also recorded the August 2011 ETS event with a slightly reduced number of stations. ~\citeA{GHO_2012} used a multibeam backprojection (MBBP) technique to detect and locate tremor. They bandpass filtered the vertical component between 5 and 9 Hz and divided the data into one-minute-long time windows. They performed beam forming of vertical component data in the frequency domain at each array to determine the slowness vectors, and backprojected the slownesses through a 3-D wavespeed model ~\cite{PRE_2003} to locate the source of the tremor for each time window. This produced 28902 tremor epicenters for one-minute-long time windows during June, 2009 - September, 2010 and 5600 epicenters during August - September, 2011.

\begin{figure}
\noindent\includegraphics[width=\textwidth, trim={0cm 2.5cm 0cm 9.5cm},clip]{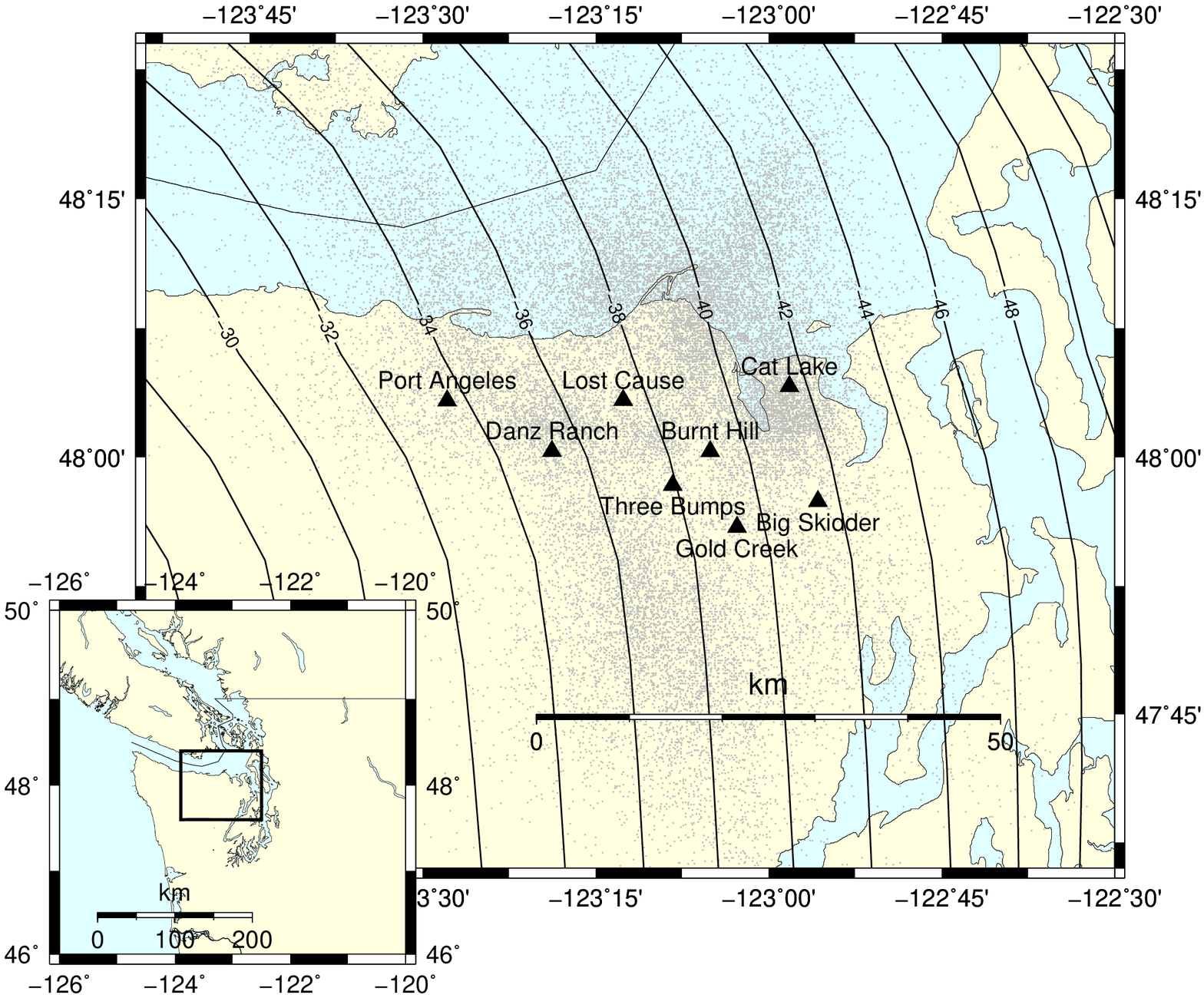}
\caption{Map showing the location of the eight arrays (black triangles) used in this study and tremor locations (grey dots) located using these arrays ~\cite{GHO_2012}. Inset shows the study area with the box marking the area covered in the main map. Contours represent a model of the depth of the plate interface ~\cite{MCC_2006}.}
\label{pngfiguresample}
\end{figure}

\section{Method}

For each array, and every 5 km by 5 km grid cell located within 25 km, we analyze the one-minute-long time windows corresponding to all the tremor epicenters located within the grid cell. For each three-component seismic station and each channel, we detrended the data, tapered the first and last 5 seconds with a Hann window, removed the instrument response, bandpass filtered between 2 and 8 Hz, and resampled the data to 20 Hz. All these preprocessing operations were done with the Python package obspy. For each seismic station and each one-minute-long time window, we cross correlated the vertical component with the East-West component and with the North-South component. Then, we stacked the cross correlation functions over all the seismic stations of the array. We call this stack $S_{i j} (\tau)$ for time lag $\tau$, the $i$-th channel ($i$=1 or 2 for the North or East channel) and the $j$-th one-minute-long time window for a given epicenter grid/array pair. At each step of this method we experimented with a linear stack, a $n$th-root stack, and a phase-weighted stack ~\cite{SCH_1997} and found that the phase-weighted stack worked best. So, all the stacks discussed here used the phase-weighted method.  Figure 2 shows an example of the envelopes of $S_{i j} (\tau)$ for the Big Skidder array for the 110 one-minute-long time windows when tremor was detected in the 5 km by 5 km grid cell located 7 kilometers southwest of the array. For all time windows, there is a peak in the envelopes of the cross-correlation $S_{i j} (\tau)$ at $\tau = 0 s$ (not shown in the figure). Additionally, for about 40\% of the time windows, we also see another peak in the envelopes of the cross-correlation $S_{i j} (\tau)$ at about $\tau = 4.5 s$. As the energy of the P-waves is expected to be higher on the vertical component, and the energy of the S-waves to be higher on the horizontal components, we interpret this peak to correspond to the time lag between the arrival of the direct P- and S-waves. This peak is only seen for the positive time lag, and no peak is seen in the negative part of the cross-correlation. \\

\begin{figure}
\noindent\includegraphics[width=\textwidth, trim={0cm 0cm 0cm 0cm},clip]{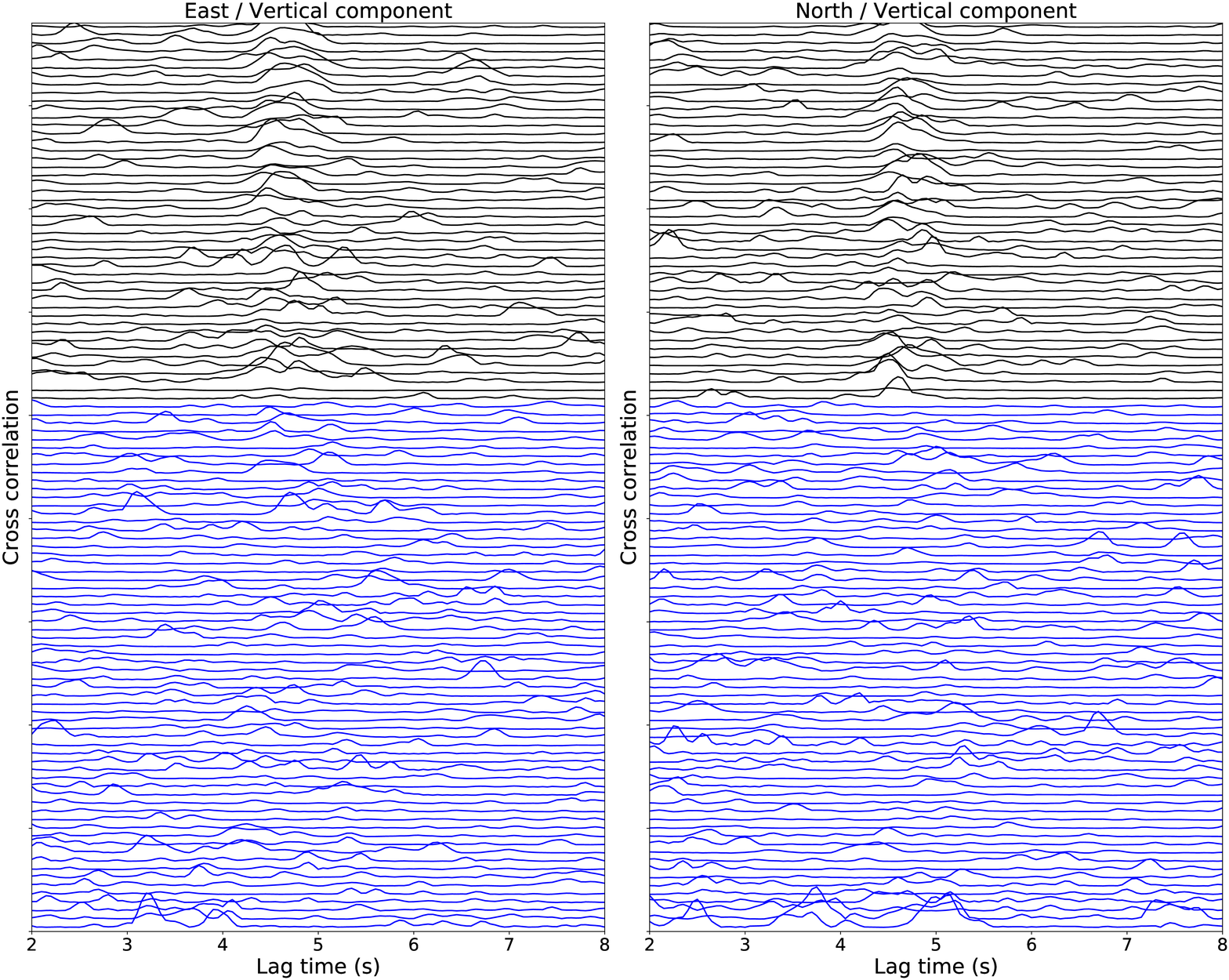}
\caption{Envelopes of the stacked cross-correlation functions for the Big Skidder array for the 110 one-minute long time windows when tremor was detected in a 5 km by 5 km grid cell located 7 kilometers southwest of the array. Top shows time windows (black) that fit the stack well. Bottom shows time windows (blue) that do not fit the stack well. Left panel is the cross-correlation of the EW component with the vertical component, and right panel is the cross-correlation of the NS component with the vertical component. The cross-correlation functions have been cut between 2 and 8 seconds to focus on the time lags around the peaks.}
\label{pngfiguresample}
\end{figure}

Only about half of the envelopes of the cross-correlation functions have a distinct peak that coincides with the peak in the stacked cross-correlation. The other cross-correlation functions show either a distinct peak at another time lag, or no clearly visible peak. This may be either because the tremor epicenter was mislocated, or because the signal-to-noise ratio is too low. \\

To determine which time window to use for further consideration we compute the theoretical value of the time lag between the P- and S-wave arrivals if the source was located on the plate boundary. We then look for the lag time corresponding to the peak of the absolute value of the stack within 1 second of the theoretical lag time. We do this nine times for each combination of 3 stacking methods acting on the stack of stations for each array and on the stacks across each of the one-minute time windows. The stacking methods are a linear stack, a $n$th-root stack or a phase-weighted stack. We define $T_{min}$ to be the minimum of these 9 lag times minus 1 s and $T_{max}$ to be the maximum of these 9 times plus 1 s. We limit the rest of our analysis to the time window $\left[ T_{min} ; T_{max} \right]$. We define $\tau_{max}$ to be the time corresponding to the peak absolute value of the phase-weighted stack within these time limits. $\tau_{max}$ turns out to also be the time corresponding to the maximum of the stacked correlations within a fixed wide window from 3-8 s for all the array / grid pairs except for one (BH, grid that is 7 km to the southeast). This suggests that there is very little, if any, bias in our method forcing our final observed S minus P times towards their theoretical values. \\

To improve the signal-to-noise ratio of the peak in the stacked cross-correlation, we divided the one-minute-long time windows into two clusters, the ones that match the stacked cross-correlations well, and the ones that do not. In order to do the clustering, we stacked the stacked cross-correlation functions, $S_{i j} (\tau)$, over all the one-minute-long time windows to obtain $S_i (\tau)$. For each one-minute time window $j$ we cross-correlated $S_i (\tau)$ with $S_{i j} (\tau)$ to obtain $S^c_{i j} (\tau)$. We want to verify whether the waveform $S_{i j} (\tau)$ around the peak $\tau_{max}$ matches well with the stack $S_i (\tau)$. For the cross-correlation between $S_{i j} (\tau)$ and $S_i (\tau)$, we keep only the values of lag times $\tau$ between $T_{min}$ and $T_{max}$. For each $i$ and $j$ we determine 3 numbers from $S^c_{i j} (\tau)$: its value at zero lag time, its maximum absolute value and the time lag at which it takes its maximum absolute value, and we determine one number from $S_{i j} (\tau)$: the ratio of its maximum absolute value $S_{i j} (\tau) _{max}$ to its rms. We considered lag times $\tau$ from 12 to 14 seconds to compute the value of the rms of the $S_{i j} (\tau)$. Each one-minute-long time window is thus associated with eight values of quality criteria, four for each of the the East and North channels. We then classified each one-minute-long time window into two different clusters, based on the value of these criteria, using a K-means clustering algorithm (function sklearn.cluster.KMeans from the Python library SciKitLearn). The K-means procedure is as follows: We choose to have 2 clusters, then we arbitrarily choose a center for each cluster. We put each one-minute-long time window into the cluster to which it is closest (based on the values of the eight criteria). Once all one-minute-long time windows have been put in a cluster, we recompute the mean of the eight criteria for each cluster, and iterate the procedure until convergence. On average, about 35\% of the time windows fit well with the stack and are kept in the first cluster, while 65\% do not fit well with the stack and are removed. For each cluster, we then stacked the envelopes of the cross-correlation functions over all the one-minute-long time windows belonging to that cluster using a phase-weighted stack. We tried using more than 2 clusters, but it did not improve the final stack of the best cluster. Figure 3 shows the stack of the envelopes of the $S_{i j} (\tau)$ over the time windows $j$ that correspond to cluster 0 (red) and cluster 1 (blue).  The clustering has greatly improved the amplitude of the peak for cluster 1, and made the peak nearly disappear for cluster 0. \\

\begin{figure}
\noindent\includegraphics[width=\textwidth, trim={0cm 0cm 0cm 0cm},clip]{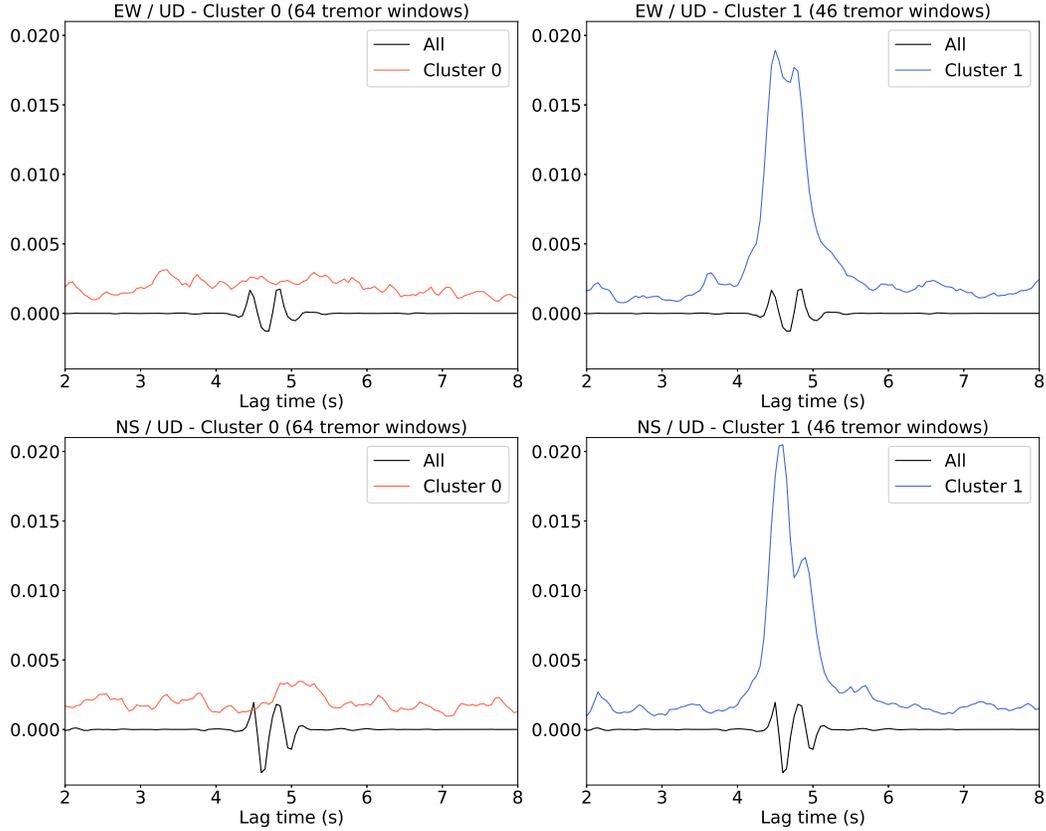}
\caption{Stack of the cross-correlation functions over all the 110 time windows from Figure 2. The black line is the stack $S_i (\tau)$ over all time windows of the stack of cross-correlation functions over all stations within the array. The red line on the left panels is the stack over all the time windows in cluster 0 of the envelopes of the stack of cross-correlation functions over stations within the array.  This corresponds to time windows that do not fit well with the stack. The blue lines on the right panels are the same as the red lines, except it uses time windows in cluster 1 (which contains the time windows that fit well with the stack). Top panels are the cross-correlation of the EW component with the vertical component, and bottom panels are the cross-correlation of the NS component with the vertical component. The stacked cross-correlation function has been cut between 2 and 8 seconds to focus on the time lags around the peak.}
\label{pngfiguresample}
\end{figure}

We did this analysis for grid cells located in a 50 km by 50 km area centered on each of the eight arrays. We thus consider up to 11 * 11 * 8 = 968 values of the time lag between the direct P-wave and S-wave. We assumed that the epicenter is at the center of the grid and determined the depth that satisfies that epicenter and the observed S-minus-P time. The velocity models are taken from the 3D velocity model from ~\citeA{MER_2020}. For each array, we looked for the closest grid point (in terms of latitude and longitude) in the 3D model by ~\citeA{MER_2020} and used the corresponding layered model of compressional and shear wave speeds to compute the tremor depth. We thus used 8 different 1D velocity models corresponding to the 8 arrays. This allows us to take into account the substantial variations of the Poisson's ratio in the East-West direction. \\

\section{Results}

To obtain robust results, we limit our analysis to grid cells where there are at least 30 one-minute-long time windows in the best cluster. We assumed that the location of the tremor source is fixed during the one-minute-long time window where we compute the cross-correlation of the seismic signal. However, during an ETS event, rapid tremor streaks have been observed to propagate up-dip or down-dip at velocities ranging on average between 30 and 110 km/h ~\cite{GHO_2010_G3}, which corresponds to a maximum source displacement of 0.9 km updip or downdip during the 30 seconds duration before and after the middle of the time window. The change in predicted S-minus-P time caused by changing source location by 0.9 km in the up- or down-dip direction is small for tremor along strike from the array or in the down dip direction.  However, the change in this lag time for tremor sources up-dip from the arrays exceeds one quarter of the dominant period of the tremor signal (period= 0.33 s) for tremor sources more than 18 km updip from an array. Thus, tremors beyond 18 km updip of an array may not add coherently during rapid tremor migrations. This method works best if the P-wave is cleanly recorded on the vertical component and the S-wave on the horizontals, so generally speaking the results are more robust for near vertical ray paths. \\

Finally, the data from some of the arrays are very noisy, which makes it hard to see a signal emerging when stacking over the one-minute-long time windows. We chose to keep only the locations for which the ratio between the maximum value of the stack of the envelopes of the cross-correlation functions to the root mean square is higher than 5. We compute the root mean square in cluster 1 for a time lag between 12 and 14 s because we do not expect any reflected wave to arrive that late after a direct wave. The corresponding stack of the envelopes of the cross-correlation functions are shown in Figure 4 for the East-West component and the North-South component. The grey dashed vertical line shows the theoretical time lag between the arrival of the direct P-wave and the arrival of the direct S-wave using the corresponding 1D velocity model for this array and the plate boundary model from ~\citeA{PRE_2003}, the grey solid line corresponds to the moment centroid of the stack. There is a good agreement between the timing of the centroid and the theoretical time lag for most of the locations of the source of the tremor. \\

\begin{figure}
\noindent\includegraphics[width=\textwidth, trim={2.5cm 0.5cm 5cm 1cm},clip]{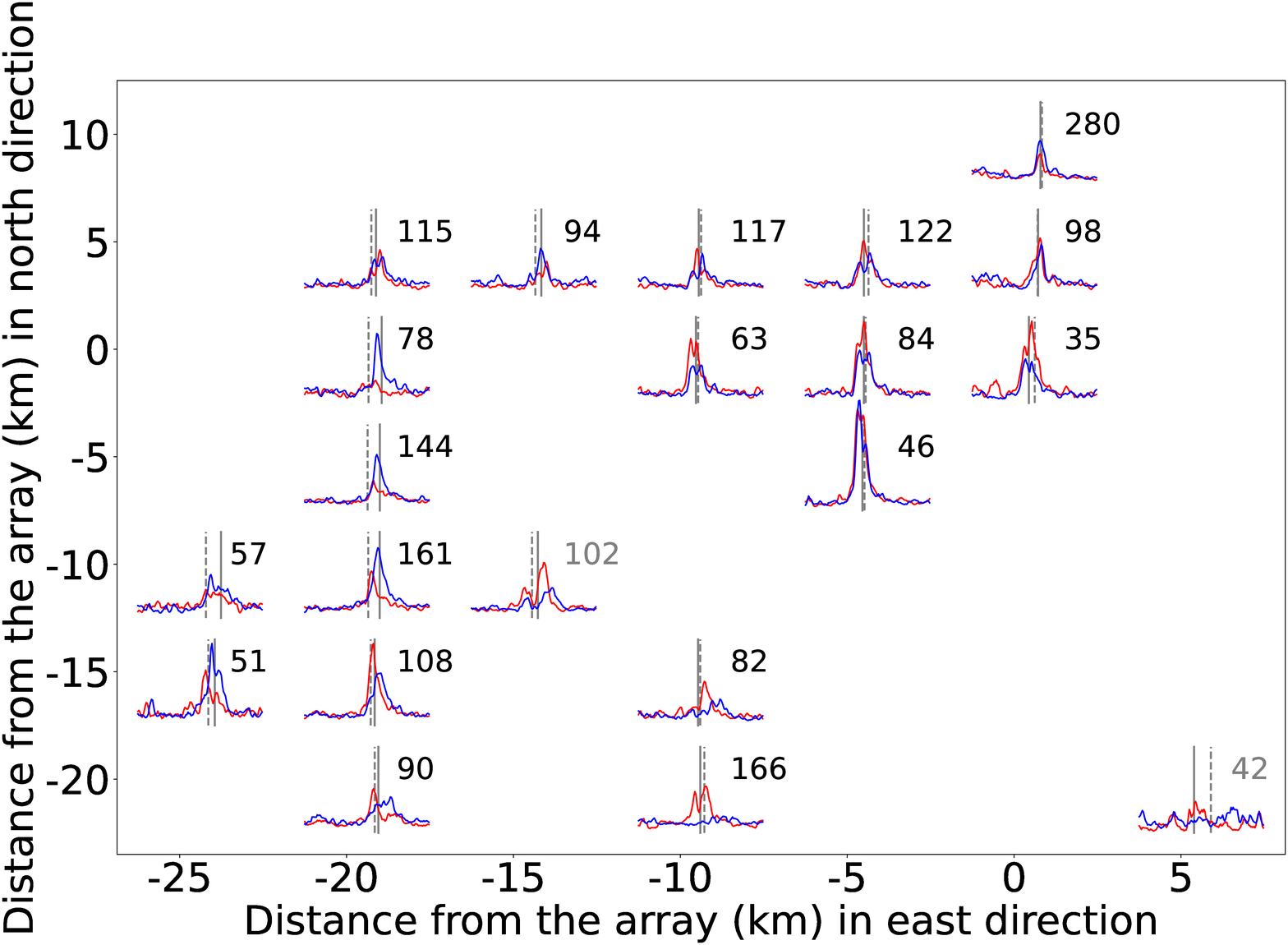}
\caption{Stack of the envelopes of the cross-correlation signals for different positions of the tremor source relative to the Big Skidder array showing the cross-correlation between the East-West and the vertical components (red lines) and between the North-South and vertical components (blue lines). The theoretical S-minus-P times (grey dashed line for the Preston model) are generally in good agreement with the centroids in cross-correlation functions (grey solid line). The location of the tremor varies from west to east (left to right) and from south to north (bottom to top). The numbers next to each graph indicate the number of one-minute-long time windows in the best cluster. For clarity, we plotted only the part of the cross-correlation signal corresponding to time lags between 2 and 8 seconds.}
\label{pngfiguresample}
\end{figure}

To compute the depth, we select the EW or the NS component with the larger maximum value of the peak. We focus on the part of the stack of the envelopes of the cross-correlation functions located between 2 seconds before and 2 seconds after the time lag $\tau_{max}$ corresponding to the peak obtained earlier. We computed the moment centroid of this part of the stack and assumed that it was equal to the time lag between the arrivals of the direct P-wave and the direct S-wave. We then computed the corresponding depth of the source of the tremor for all the locations of the source of the tremor. To verify the effect of the width of the time window for which we compute the moment centroid, we compute the centroid and the corresponding depth of the source for widths equal to 2 seconds, 4 seconds, and 6 seconds. We compute the variations in depths when decreasing the width from 4 seconds to 2 seconds. 87\% of the differences in depth are under 2 kilometers. The median difference is 0.064 km, so there is no bias towards shallower or deeper depths when we decrease the width of the time window. When we compute the variations in depths when increasing the width from 4 seconds to 6 seconds, 84\% of the differences in depth are under 2 kilometers and the median difference is 0.054 km. The corresponding stacks of the envelopes are shown in Figure 4 and Figures S1 to S7 in the supplementary material. For the two arrays Cat Lake and Lost Cause, there are very few source-array locations that have both enough tremor and a high ratio between the peak and the root mean square. Moreover, for the locations where we have a peak, the peak is often not very clear and stretched along the time axis. We did not use data from these two arrays. For the Port Angeles array, there is only a clear peak for a near vertical incidence of the seismic waves. We choose to keep this array in the analysis in order to have two additional data points in the westernmost region of the study area. \\

To estimate the uncertainty on the depth of the source of the tremor, we computed the width of the stack of the envelopes of the cross-correlation functions at half the maximum of the peak. We then computed the associated depth difference using the 1D velocity model. As we computed the envelopes of the cross-correlation functions before stacking, the peak is large and also the associated uncertainty on the depth. Nevertheless, about 75\% of the data points have an uncertainty under 8km (Figure 5). The uncertainties are higher for the Burnt Hill array, and are almost always lower than 8km for the five other arrays. We tried different stacking methods, both for the stacking of the correlation functions over all the seismic stations of a given array, and for the stacking of the envelopes of those stacks over the one-minute-long time windows when tremor is detected. We experimented with a linear stack, a $n$th-root stack, and a phase-weighted stack, and found that using a phase-weighted stack for both the stacking over the stations and the stacking over one-minute-long time windows gave the lower uncertainty on the tremor depth. \\

\begin{figure}
\noindent\includegraphics[width=\textwidth, trim={1cm 5cm 3.5cm 4cm},clip]{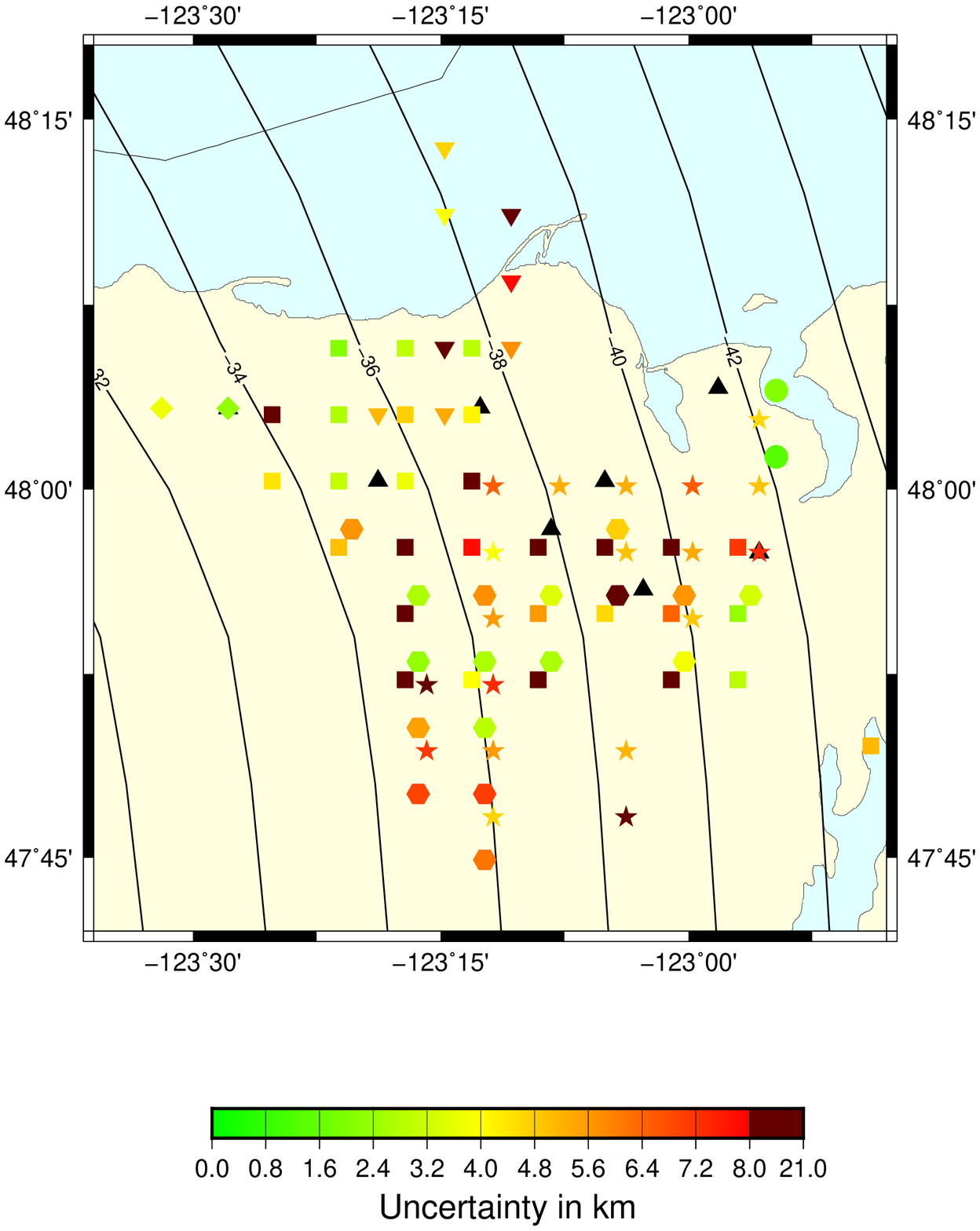}
\caption{Map of the uncertainty on the depth of the tremor based on the width of the stack of the envelopes of the cross-correlation functions for arrays Burnt Hill (squares), Big Skidder (stars), Danz Ranch (inverted triangles),  Gold Creek (large circles), Port Angeles (diamonds), and Three Bumps (hexagons).}
\label{pngfiguresample}
\end{figure}

In the following, we kept only the data points for which the maximum amplitude of the stack is higher than 0.05. Figures 6 and 7 show respectively a map and three cross-sections of the depth of the source of the tremor. Figure 8 shows a map of the distance between the source of the tremor and the plate boundary from the Preston model, alongside with the depth of the low-frequency earthquake families observed by ~\citeA{SWE_2019} and ~\citeA{CHE_2017_JGR,CHE_2017_G3}. There is a good agreement between the depth of the source of the tremor, the depth of the low-frequency earthquake families, and the depth of the plate boundary. On the cross-section figures, we compare the distance of the source of the tremor to the plate boundary for different locations of the cross-section. We note that the depth of the tremor matches well the depth of the low-frequency earthquakes. \\

\begin{figure}
\noindent\includegraphics[width=\textwidth, trim={1cm 5cm 3.5cm 4cm},clip]{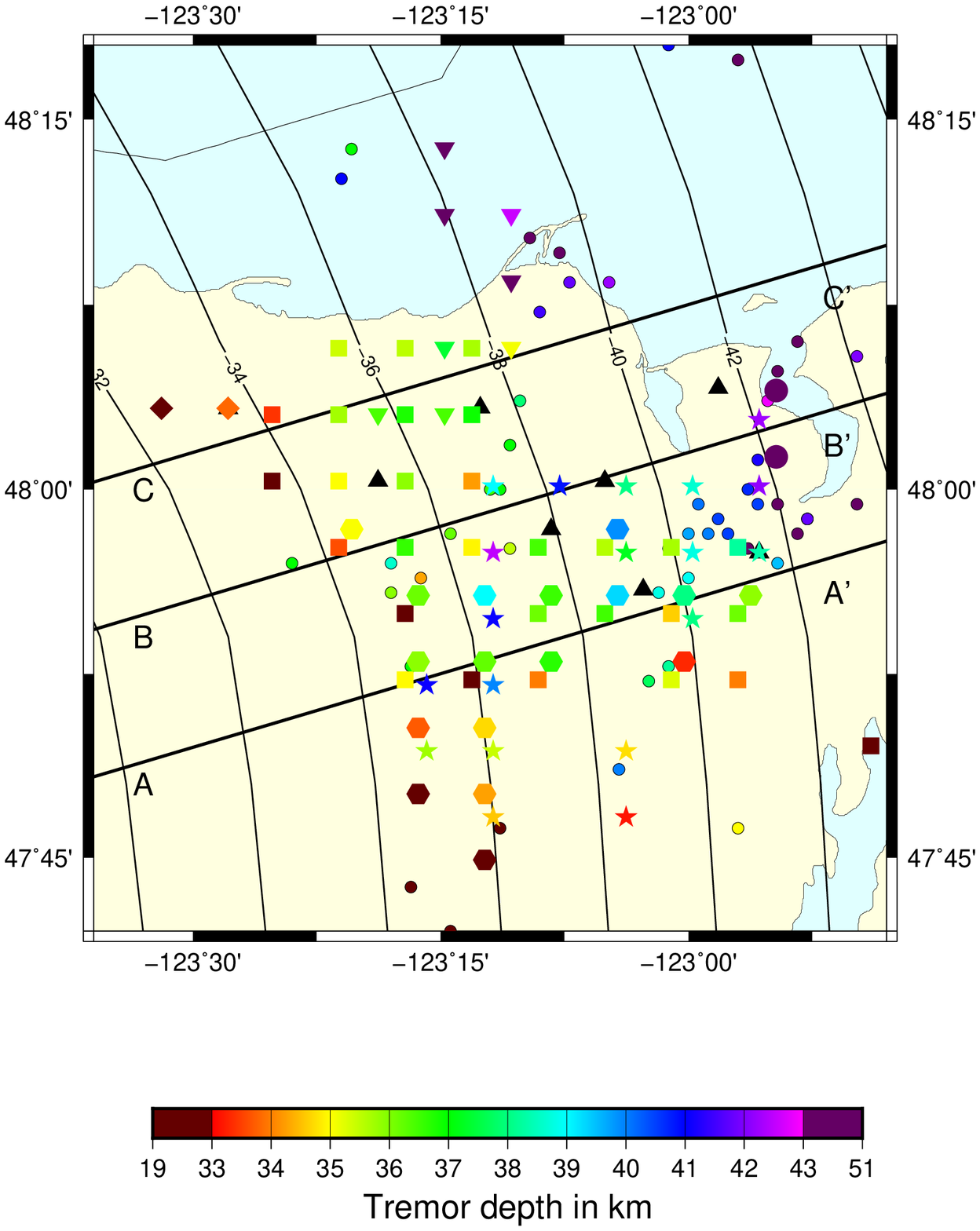}
\caption{Map of the depth of the source of the tremor and of the low-frequency earthquake families (small filled circles) identified by ~\citeA{SWE_2019} and ~\citeA{CHE_2017_JGR,CHE_2017_G3}. The three black lines indicate the positions of the cross-sections shown in Figure 7. Tremor depths are from arrays Burnt Hill (squares), Big Skidder (stars), Danz Ranch (inverted triangles),  Gold Creek (large circles), Port Angeles (diamonds), and Three Bumps (hexagons)}
\label{pngfiguresample}
\end{figure}

\begin{figure}
\noindent\includegraphics[width=5cm, trim={4.5cm 0.5cm 6cm 0cm},clip, angle=270]{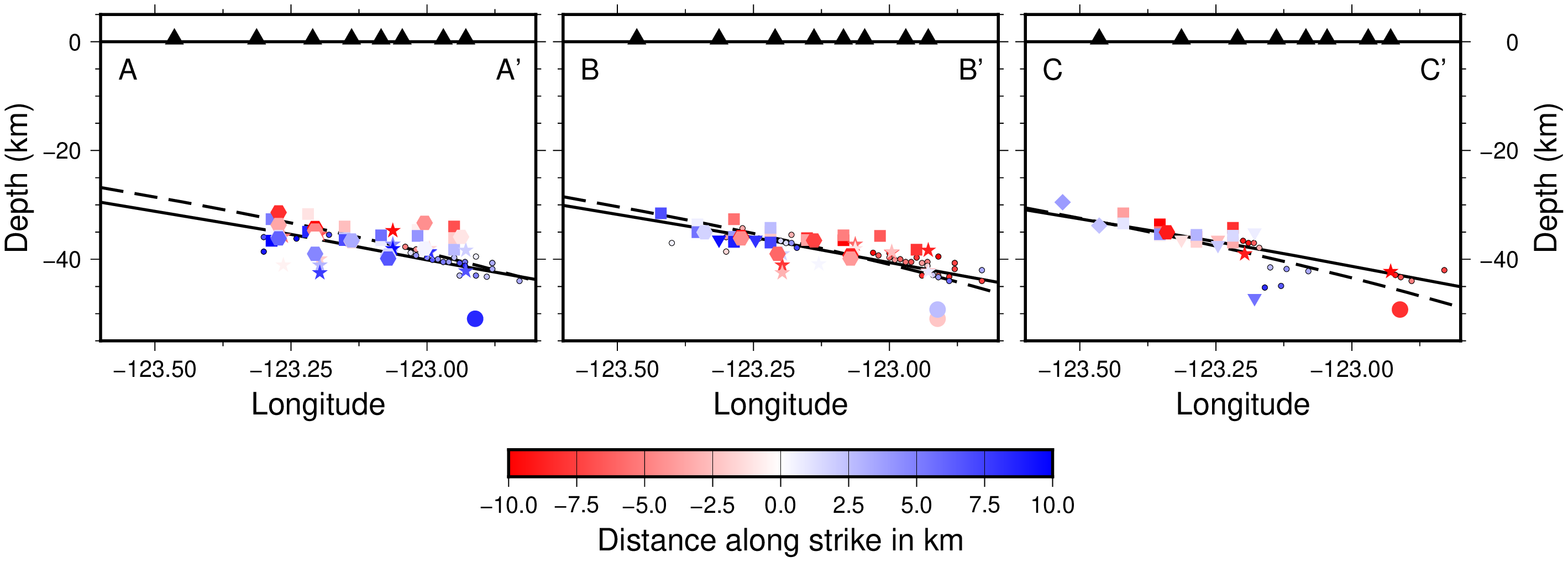}
\caption{Cross-sections showing depth of the tremor and of the low-frequency earthquake families (small filled circles) identified by ~\citeA{SWE_2019} and ~\citeA{CHE_2017_JGR,CHE_2017_G3} for cross sections A-A' , B-B', and C-C' shown in Figure 6. The black line corresponds to the plate boundary profile of ~\citeA{MCC_2006} along the southern most black line shown on Figure 6, and the dashed line to the plate boundary profile of ~\citeA{PRE_2003}. The color bar shows the distance of the tremor and the low-frequency earthquakes along the strike. Only tremor and low-frequency earthquakes less than 10 km away from the profile line are shown. Tremor depths are from arrays Burnt Hill (squares), Big Skidder (stars), Danz Ranch (inverted triangles),  Gold Creek (large circles), Port Angeles (diamonds), and Three Bumps (hexagons).}
\label{pngfiguresample}
\end{figure}

\begin{figure}
\noindent\includegraphics[width=\textwidth, trim={1cm 5cm 3.5cm 4cm},clip]{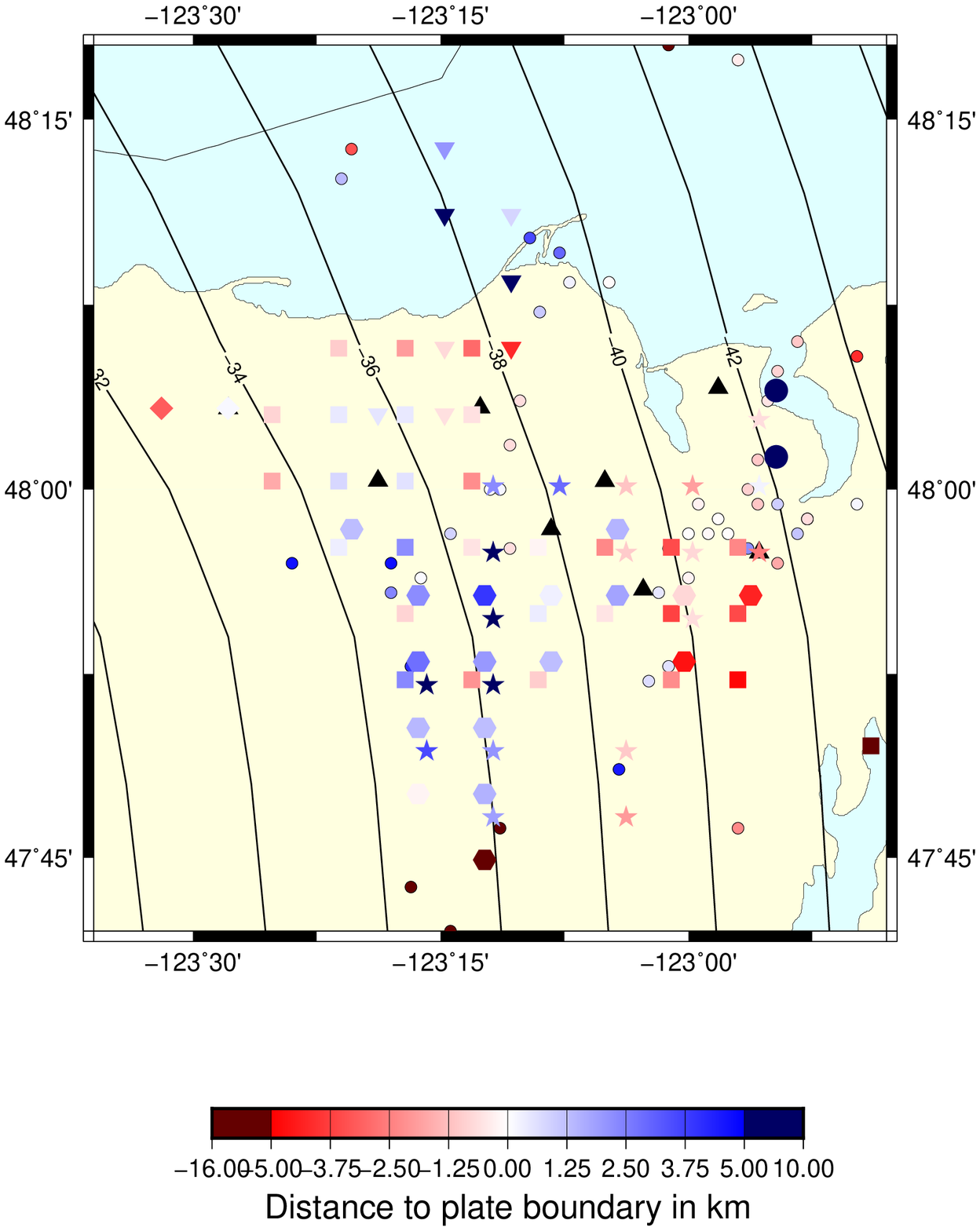}
\caption{Map of the distance between the plate boundary (from the Preston model) of the tremor and of the low-frequency earthquake families (filled circles) identified by ~\citeA{SWE_2019} and ~\citeA{CHE_2017_JGR,CHE_2017_G3}). Tremor located below the plate boundary is blue, and above is red. Tremor depths are from arrays Burnt Hill (squares), Big Skidder (stars), Danz Ranch (inverted triangles),  Gold Creek (large circles), Port Angeles (diamonds), and Three Bumps (hexagons).}
\label{pngfiguresample}
\end{figure}

\section{Discussion}

Previous studies have shown evidence for seismic anisotropy near the slab interface ~\cite{NIK_2009} and in the overriding continental crust ~\cite{CAS_1996}. If the source of the tremor is located in or under the anisotropic layer, the time lag between the arrival of the direct P-wave and the arrival of the direct S-wave could be different depending on whether we consider the East-West component or the North-South component. To verify whether possible anisotropy could affect our results, we computed the time difference between the timing of the maximum of the stack of the envelopes of the stacked cross-correlation functions between the East-West component and the vertical component, and between the North-South component and the vertical component. We also computed the associated difference in depth of the source of the tremor (see Figure 9) for all arrays and all locations of the tremor. The difference in time is generally less than 0.25 s while the difference in depth is generally less than 2 kilometers, which is similar to the uncertainty on the tremor depth. As the time difference between the East-West component and the North-South component due to anisotropy is expected to vary depending on the relative position of the array compared to the source of the tremor, we also plotted the time difference as a function of the distance from source to array, and as a function of the azimuth, as well as the corresponding depth difference (Figure 9). The time difference does not increase with the distance from the source to the array, and there is no seismic path orientation that gives bigger time difference. Thus, seismic anisotropy does not seem to have a significant effect on the time lags between the arrival of the direct P-wave and the arrival of the direct S-wave, and can be neglected. ~\citeA{NIK_2009} observed anisotropy in the low-velocity layer beneath station GNW, located in the eastern Olympic Peninsula, south of the eight seismic arrays used in this study. If the tremor is located above the low-velocity layer, we do not expect this source of seismic anisotropy to introduce a significant effect on our measured time lags. ~\citeA{NIK_2009} locate the low-velocity layer above the plate boundary, in the lower continental crust, which is shallower than the location given by ~\citeA{BOS_2013}, who locates the low-velocity layer in the upper oceanic crust. However, even if a low-velocity layer with 5 \% anisotropy is located in the lower continental crust as indicated by ~\citeA{NIK_2009}, the resulting difference in time lags should not be more than a few tenths of a second, and the resulting difference in depth should not be more than 2 kilometers. ~\citeA{CAS_1996} observed anisotropy in the continental crust, especially in the upper 20 kilometers to the north of our arrays. However, the anisotropy resulted in time delays of the seismic waves of no more than 0.32s for deep earthquakes (40-60 km depth) and not more than 0.20s for shallow earthquakes (15-30 km depth). The corresponding difference in tremor depth is similar to the uncertainty. \\

\begin{figure}
\noindent\includegraphics[width=\textwidth, trim={0cm 0cm 0cm 0cm},clip]{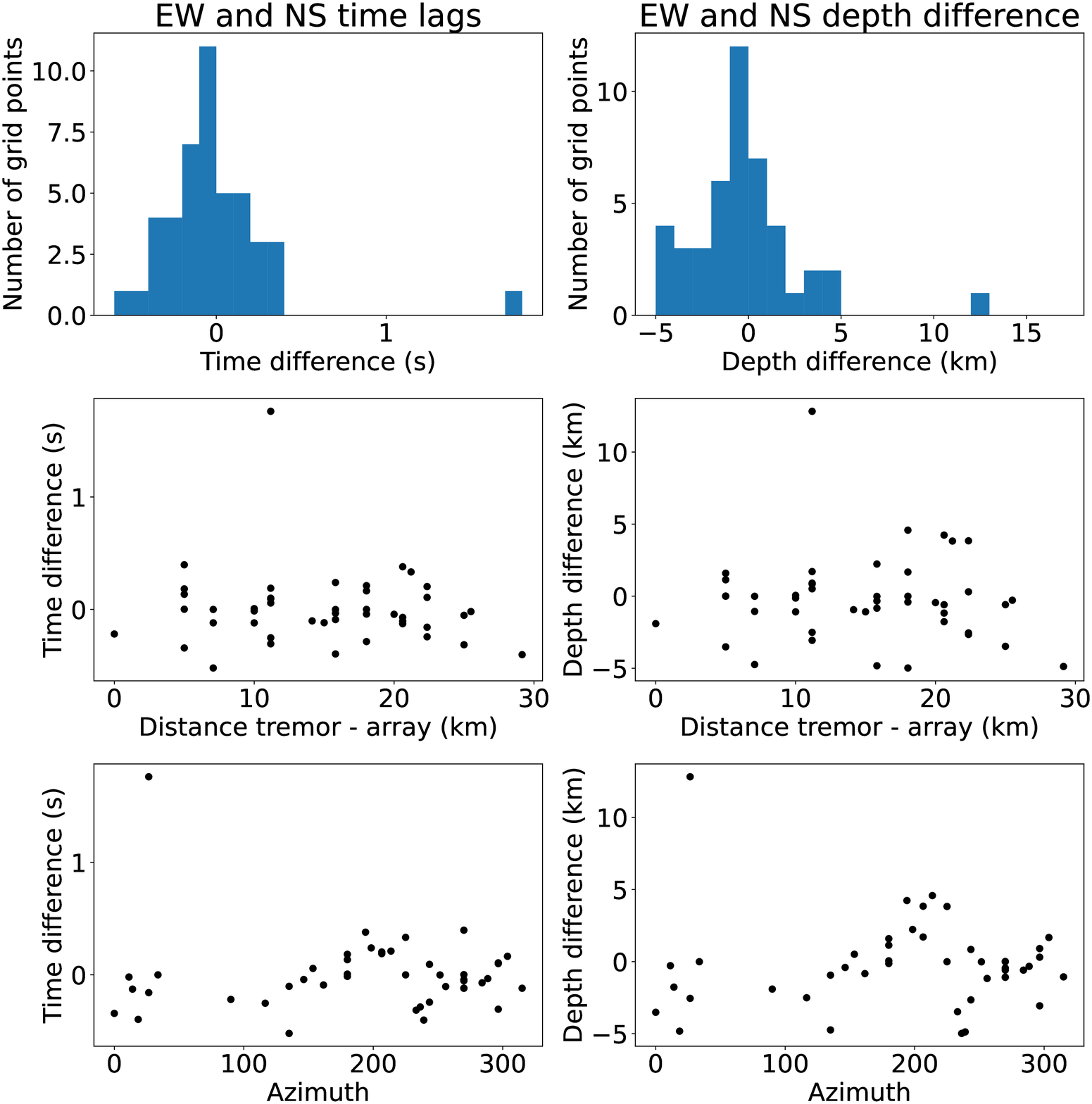}
\caption{Left: S-minus-P times measured from envelopes of cross-correlation functions on the East-West component minus those from the North-South component. Right: Corresponding difference in inferred tremor depths.  These are plotted as histograms (top), versus epicentral distance between tremors and arrays (middle) and tremor to array azimuth. There is no systematic signal that might be caused by anisotropy.}
\label{pngfiguresample}
\end{figure}

Several approximations have been made to compute the tremor depth. First, we stacked over all the tremor in 5km by 5km grid cell. The depth difference corresponding to epicentral tremor locations at the center versus the edge of a cell varies from about 1 to 2 km for cells near the array to cells 25 km from an array. Second, the Vp / Vs ratio is not very well known. A variation of about 1\% of the Vp / Vs ratio used to compute the depth  would lead to a variation of about 1-1.5 km of the corresponding tremor depth (Figure 10). \\

\begin{figure}
\noindent\includegraphics[width=\textwidth, trim={0cm 0cm 0cm 0cm},clip]{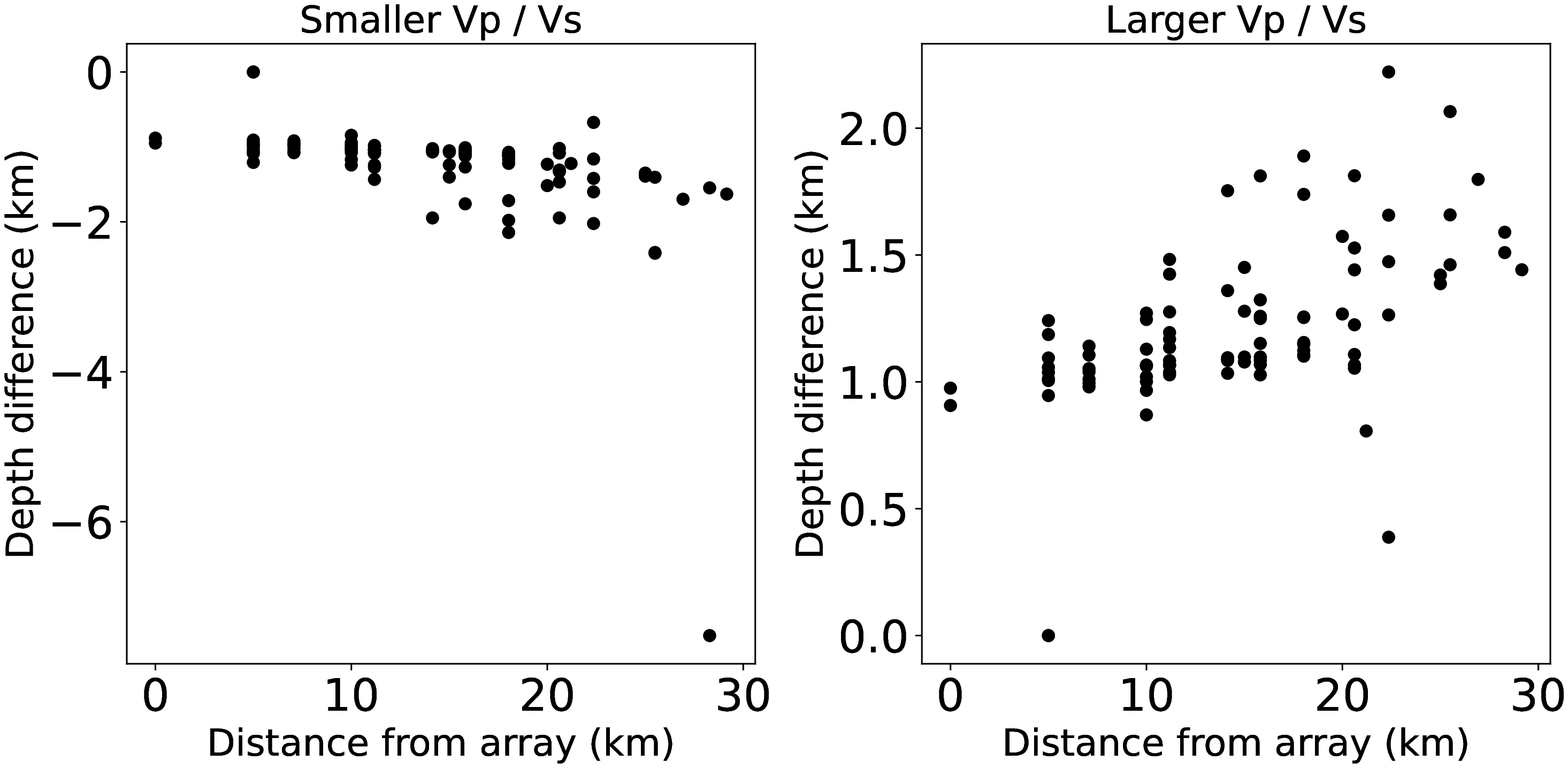}
\caption{Variations of the depth of the tremor and we decrease the Vp / Vs ratio of 1\% (left) and when we increase the Vp / Vs ratio of 1\% (right). A change in the Vp / Vs ratio corresponds to a change in the theoretical time lag between direct P-wave and direct S-wave if the source is located on the plate boundary. Thus, it changes the time interval over which we compute the centroid of the stacked envelopes, and thus the time lag we use to compute the depth.}
\label{pngfiguresample}
\end{figure}

As proposed by ~\citeA{KAO_2009}, the source of the tremor could be distributed inside a layer that is 15 or more kilometers thick. We have assumed that all the tremor within a given grid cell originate from the same depth, and averaged over all the data to get the tremor depth. However, instead of being located on the same plane near the plate boundary, the tremor may be scattered over a layer surrounding the plate boundary. To compute the thickness of this layer, for each location of the array and the source of the tremor, we computed for each one-minute-long time window for which the cross-correlation function matches well the stacked cross-correlation the time lag between the time corresponding to the maximum absolute value for the cross-correlation function and the time corresponding to the maximum absolute value for the stacked cross-correlation. We thus obtained a distribution of time lags and the corresponding distribution of depths, and we tried to estimate the scale of the interval over which the depths vary. An example of time lags distribution is shown in Figure S8 of the supplementary material. Using the standard deviation could lead to overestimate the width of the interval, and the corresponding thickness of the tremor layer, as the standard deviation is very sensitive to outliers. Instead, we use the Qn estimator of ~\citeA{ROU_1993}, which is a more robust estimator of scale, similar to the median absolute deviation (MAD). Qn is equal to the $k$th order statistic of the $\begin{pmatrix} n \\ 2 \end{pmatrix}$ interpoint distances where $n$ is the number of points, and $k = \begin{pmatrix} h \\ 2 \end{pmatrix}$ with $h = \left[ \frac{n}{2} \right] + 1$ and $\left[ x \right]$ denote the integer part of $x$. As the MAD, the Qn estimator is not very sensitive to outliers, and it has the advantage of being suitable for asymmetric distributions, contrary to the MAD, which attaches equal importance to positive and negative deviations from the median. We then used the scale of the interval over which the tremor depth varies as an estimate of the thickness of the tremor layer (see Figure 11). The average thickness of the tremor zone is about 1.6 km, while nearly all the values of the thickness are less than 3km, which is substantially lower than the depth extent from ~\citeA{KAO_2009}. The smallest values for the thickness are about 0.8-2 kilometers. \\

\begin{figure}
\noindent\includegraphics[width=\textwidth, trim={1cm 5cm 3.5cm 4cm},clip]{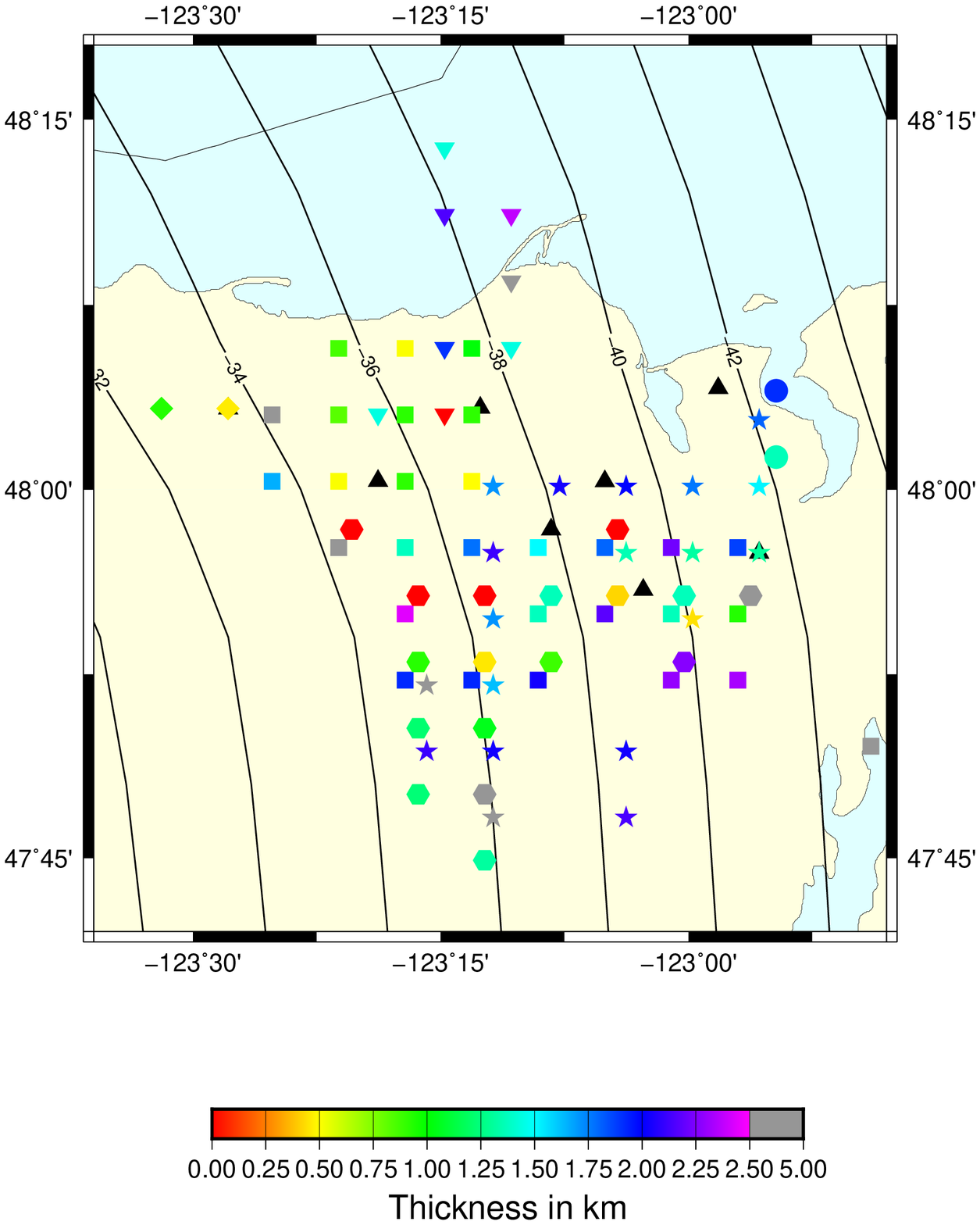}
\caption{Map of the thickness of the tremor layer estimated from scatter of individual 1-minute tremor windows. Tremor widths are from arrays Burnt Hill (squares), Big Skidder (stars), Danz Ranch (inverted triangles),  Gold Creek (large circles), Port Angeles (diamonds), and Three Bumps (hexagons).}
\label{pngfiguresample}
\end{figure}

Another way of estimating the thickness of the tremor zone is to fit a plane with a linear regression from all the values of the depth and compute the residuals from the regression (Figure 12). The Qn estimator of scale ~\cite{ROU_1993} for the error between the tremor depth and the fitted plane is only 1.3 km. \\

\begin{figure}
\noindent\includegraphics[width=\textwidth, trim={1cm 5cm 3.5cm 4cm},clip]{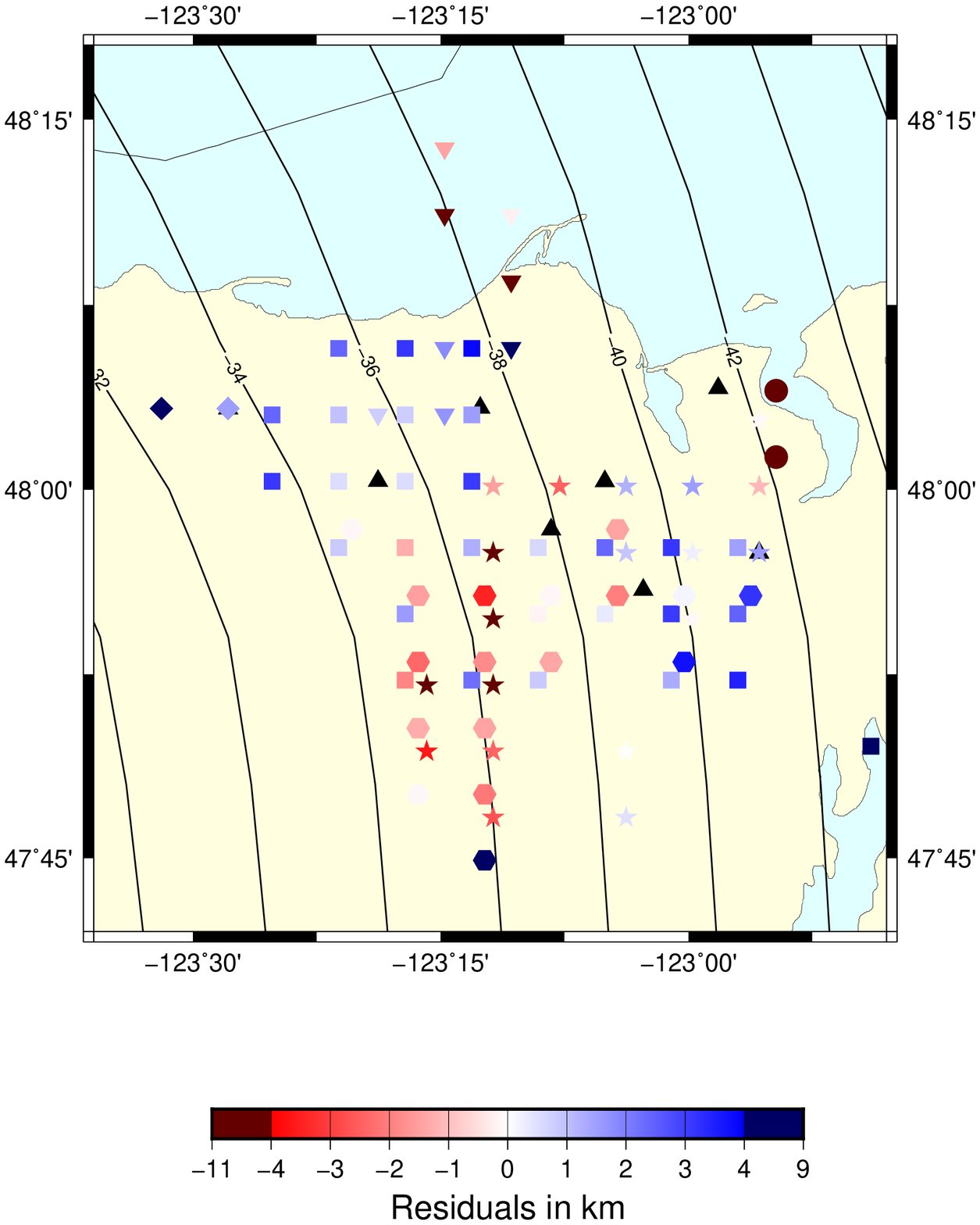}
\caption{Map of the residuals from the linear regression fitting all the depths to a plane. Residuals are from arrays Burnt Hill (squares), Big Skidder (stars), Danz Ranch (inverted triangles),  Gold Creek (large circles), Port Angeles (diamonds), and Three Bumps (hexagons).}
\label{pngfiguresample}
\end{figure}

In Table 1, we report the values of the median, the median absolute deviation (MAD), and the Qn estimator of scale for the difference between the depth of the tremor and the low-frequency earthquakes and the depth of the plate boundary, for the McCrory model and for the Preston model. Small negative values of the median correspond to tremor and low-frequency earthquakes located a little bit above the plate boundary. The depth range of the tremor (Qn = 1.3 km) is larger than the depth range of the low-frequency earthquake families from ~\citeA{CHE_2017_JGR,CHE_2017_G3} (Qn = 0.9 km). It is a also larger than the thickness of the flow channel where ~\citeA{SAM_2021} suggest that the LFEs are generated (0.5 to 1.5 km). However, this may be due to the uncertainty on the determination of the tremor depth. The tremor and low-frequency earthquakes are slightly closer to the plate boundary for the Preston model than for the McCrory model. This may be because the P-wave velocity model of ~\citeA{PRE_2003} was used as the initial velocity model by ~\citeA{MER_2020} to compute the full P- and S-wave velocity model that we use to get tremor depth from S minus P times. The Preston P-wave velocity model was also used by ~\citeA{CHE_2017_JGR,CHE_2017_G3} to locate the low-frequency earthquake families. \\

\begin{table}
\caption{Summary of distances to plate boundary}
\centering
\begin{tabular}{l c c c}
\hline
 & Median & MAD & Qn \\ 
\hline
Tremor depth - Preston depth & - 0.207 km & 2.623 km & 1.313 km \\
Tremor depth - McCrory depth & - 0.850 km & 2.429 km & 1.390 km \\
LFE families depth - Preston depth & - 0.189 km & 1.312 km & 0.871 km \\
LFE families depth - McCrory depth & - 0.495 km & 1.805 km & 0.988 km \\
\hline
\end{tabular}
\end{table}

\section{Conclusion}

We developed a method to estimate the depth of the source of the tectonic tremor, and the depth extent of the region from which the tremor originates, using S minus P times determined from lag times of stacked cross-correlations of horizontal and vertical components of seismic recordings from small aperture arrays in the Olympic Peninsula, Washington. We found that the source of the tremor is located close to the plate boundary in a region no more than 2-3 kilometers thick. The source of the tremor is thus distributed over a slightly wider depth range than the low-frequency earthquakes. However, due to the depth uncertainty, it is difficult to conclude whether the tremor is located near the top  of the subducting oceanic crust, in the lower continental crust just above the plate boundary, in a layer distributed above and below the plate boundary, or confined to a very narrow plate boundary. The location of the tremor relative to the low-velocity layer also observed near the plate boundary also remains uncertain.  Tremor and LFE depths are consistent with filling a volume in the upper subducted crust that is characterized by high fluid pressure and very low S-wave velocities described as the preferred model by ~\citeA{BOS_2013}. \\

\acknowledgments
The authors would like to thank A. Ghosh for sharing his tremor catalog. We also would like to thank Charles Sammis and an anonymous reviewer whose reviews have helped improve the paper. This project was funded by NSF grant EAR-1358512. A.D. would like to thank the Integral Environmental Big Data Research Fund from the College of the Environment of University of Washington, which funded cloud computing resources to carry out the data analyses. The seismic recordings used for this analysis can be downloaded from the IRIS website using network code XG, 2009-2011. Most of the figures were done using GMT ~\cite{WES_1991}. The Python scripts used to analyze the data and make the figures can be found on the first author's Github account, accessible through Zenodo ~\cite{ariane_ducellier_2021_5725990}.


%
%

\bibliography{bibliography}

\begin{thebibliography}{}

\bibitem [\protect \citeauthoryear {%
Audet%
, Bostock%
, Christensen%
\BCBL {}\ \BBA {} Peacock%
}{%
Audet%
\ \protect \BOthers {.}}{%
{\protect \APACyear {2009}}%
}]{%
AUD_2009}
\APACinsertmetastar {%
AUD_2009}%
\begin{APACrefauthors}%
Audet, P.%
, Bostock, M.%
, Christensen, N.%
\BCBL {}\ \BBA {} Peacock, S.%
\end{APACrefauthors}%
\unskip\
\newblock
\APACrefYearMonthDay{2009}{}{}.
\newblock
{\BBOQ}\APACrefatitle {Seismic evidence for overpressured subducted oceanic
  crust and megathrust fault sealing} {Seismic evidence for overpressured
  subducted oceanic crust and megathrust fault sealing}.{\BBCQ}
\newblock
\APACjournalVolNumPages{Nature}{457}{}{76-78}.
\PrintBackRefs{\CurrentBib}

\bibitem [\protect \citeauthoryear {%
Audet%
\ \BBA {} B\"urgmann%
}{%
Audet%
\ \BBA {} B\"urgmann%
}{%
{\protect \APACyear {2014}}%
}]{%
AUD_2014}
\APACinsertmetastar {%
AUD_2014}%
\begin{APACrefauthors}%
Audet, P.%
\BCBT {}\ \BBA {} B\"urgmann, R.%
\end{APACrefauthors}%
\unskip\
\newblock
\APACrefYearMonthDay{2014}{}{}.
\newblock
{\BBOQ}\APACrefatitle {Possible control of subduction zone slow-earthquake
  periodicity by silica enrichment} {Possible control of subduction zone
  slow-earthquake periodicity by silica enrichment}.{\BBCQ}
\newblock
\APACjournalVolNumPages{Nature}{510}{}{389-393}.
\PrintBackRefs{\CurrentBib}

\bibitem [\protect \citeauthoryear {%
Bostock%
}{%
Bostock%
}{%
{\protect \APACyear {2013}}%
}]{%
BOS_2013}
\APACinsertmetastar {%
BOS_2013}%
\begin{APACrefauthors}%
Bostock, M.%
\end{APACrefauthors}%
\unskip\
\newblock
\APACrefYearMonthDay{2013}{}{}.
\newblock
{\BBOQ}\APACrefatitle {The {Moho} in subduction zones} {The {Moho} in
  subduction zones}.{\BBCQ}
\newblock
\APACjournalVolNumPages{Tectonophysics}{609}{}{547-557}.
\PrintBackRefs{\CurrentBib}

\bibitem [\protect \citeauthoryear {%
Bostock%
, Royer%
, Hearn%
\BCBL {}\ \BBA {} Peacock%
}{%
Bostock%
\ \protect \BOthers {.}}{%
{\protect \APACyear {2012}}%
}]{%
BOS_2012}
\APACinsertmetastar {%
BOS_2012}%
\begin{APACrefauthors}%
Bostock, M.%
, Royer, A.%
, Hearn, E.%
\BCBL {}\ \BBA {} Peacock, S.%
\end{APACrefauthors}%
\unskip\
\newblock
\APACrefYearMonthDay{2012}{}{}.
\newblock
{\BBOQ}\APACrefatitle {Low frequency earthquakes below southern {Vancouver}
  {Island}} {Low frequency earthquakes below southern {Vancouver}
  {Island}}.{\BBCQ}
\newblock
\APACjournalVolNumPages{Geochemistry Geophysics Geosystems}{13}{}{Q11007}.
\PrintBackRefs{\CurrentBib}

\bibitem [\protect \citeauthoryear {%
Cassidy%
\ \BBA {} Bostock%
}{%
Cassidy%
\ \BBA {} Bostock%
}{%
{\protect \APACyear {1996}}%
}]{%
CAS_1996}
\APACinsertmetastar {%
CAS_1996}%
\begin{APACrefauthors}%
Cassidy, J.%
\BCBT {}\ \BBA {} Bostock, M.%
\end{APACrefauthors}%
\unskip\
\newblock
\APACrefYearMonthDay{1996}{}{}.
\newblock
{\BBOQ}\APACrefatitle {Shear-wave splitting above the subducting {Juan} de
  {Fuca} plate} {Shear-wave splitting above the subducting {Juan} de {Fuca}
  plate}.{\BBCQ}
\newblock
\APACjournalVolNumPages{Geophysical Research Letters}{23}{9}{941-944}.
\PrintBackRefs{\CurrentBib}

\bibitem [\protect \citeauthoryear {%
Chestler%
\ \BBA {} Creager%
}{%
Chestler%
\ \BBA {} Creager%
}{%
{\protect \APACyear {2017a}}%
}]{%
CHE_2017_JGR}
\APACinsertmetastar {%
CHE_2017_JGR}%
\begin{APACrefauthors}%
Chestler, S.%
\BCBT {}\ \BBA {} Creager, K.%
\end{APACrefauthors}%
\unskip\
\newblock
\APACrefYearMonthDay{2017a}{}{}.
\newblock
{\BBOQ}\APACrefatitle {Evidence for a scale-limited low-frequecny earthquake
  source process} {Evidence for a scale-limited low-frequecny earthquake source
  process}.{\BBCQ}
\newblock
\APACjournalVolNumPages{Journal of Geophysical Research. Solid
  Earth}{122}{}{3099-3114}.
\newblock
\APACrefnote{doi:10.1002/2016jb013717}
\PrintBackRefs{\CurrentBib}

\bibitem [\protect \citeauthoryear {%
Chestler%
\ \BBA {} Creager%
}{%
Chestler%
\ \BBA {} Creager%
}{%
{\protect \APACyear {2017b}}%
}]{%
CHE_2017_G3}
\APACinsertmetastar {%
CHE_2017_G3}%
\begin{APACrefauthors}%
Chestler, S.%
\BCBT {}\ \BBA {} Creager, K.%
\end{APACrefauthors}%
\unskip\
\newblock
\APACrefYearMonthDay{2017b}{}{}.
\newblock
{\BBOQ}\APACrefatitle {A model for low-frequency earthquake slip} {A model for
  low-frequency earthquake slip}.{\BBCQ}
\newblock
\APACjournalVolNumPages{Geochemistry, Geophysics, Geosystems}{18}{}{4690-4708}.
\newblock
\APACrefnote{doi:10.1002/2017gc007253}
\PrintBackRefs{\CurrentBib}

\bibitem [\protect \citeauthoryear {%
Ducellier%
}{%
Ducellier%
}{%
{\protect \APACyear {2021}}%
}]{%
ariane_ducellier_2021_5725990}
\APACinsertmetastar {%
ariane_ducellier_2021_5725990}%
\begin{APACrefauthors}%
Ducellier, A.%
\end{APACrefauthors}%
\unskip\
\newblock
\APACrefYearMonthDay{2021}{{\APACmonth{11}}}{}.
\newblock
\APACrefbtitle {ArianeDucellier/timelags: Revised manuscript.}
  {Arianeducellier/timelags: Revised manuscript.}
\newblock
\APACaddressPublisher{}{Zenodo}.
\newblock
\begin{APACrefURL} \url{https://doi.org/10.5281/zenodo.5725990}
  \end{APACrefURL}
\newblock
\begin{APACrefDOI} \doi{10.5281/zenodo.5725990} \end{APACrefDOI}
\PrintBackRefs{\CurrentBib}

\bibitem [\protect \citeauthoryear {%
Fagereng%
\ \BBA {} Diener%
}{%
Fagereng%
\ \BBA {} Diener%
}{%
{\protect \APACyear {2011}}%
}]{%
FAG_2011}
\APACinsertmetastar {%
FAG_2011}%
\begin{APACrefauthors}%
Fagereng, {\r A}.%
\BCBT {}\ \BBA {} Diener, J.%
\end{APACrefauthors}%
\unskip\
\newblock
\APACrefYearMonthDay{2011}{}{}.
\newblock
{\BBOQ}\APACrefatitle {Non‐volcanic tremor and discontinuous slab
  dehydration} {Non‐volcanic tremor and discontinuous slab
  dehydration}.{\BBCQ}
\newblock
\APACjournalVolNumPages{Geophysical Research Letters}{38}{}{L15302}.
\PrintBackRefs{\CurrentBib}

\bibitem [\protect \citeauthoryear {%
Fagereng%
, G.W.B.%
\BCBL {}\ \BBA {} Diener%
}{%
Fagereng%
\ \protect \BOthers {.}}{%
{\protect \APACyear {2014}}%
}]{%
FAG_2014}
\APACinsertmetastar {%
FAG_2014}%
\begin{APACrefauthors}%
Fagereng, {\r A}.%
, G.W.B., H.%
\BCBL {}\ \BBA {} Diener, J.%
\end{APACrefauthors}%
\unskip\
\newblock
\APACrefYearMonthDay{2014}{}{}.
\newblock
{\BBOQ}\APACrefatitle {Brittle-viscous deformation, slow slip, and tremor}
  {Brittle-viscous deformation, slow slip, and tremor}.{\BBCQ}
\newblock
\APACjournalVolNumPages{Geophysical Research Letters}{41}{12}{4159-4167}.
\PrintBackRefs{\CurrentBib}

\bibitem [\protect \citeauthoryear {%
Ghosh%
, Vidale%
\BCBL {}\ \BBA {} Creager%
}{%
Ghosh%
\ \protect \BOthers {.}}{%
{\protect \APACyear {2012}}%
}]{%
GHO_2012}
\APACinsertmetastar {%
GHO_2012}%
\begin{APACrefauthors}%
Ghosh, A.%
, Vidale, J.%
\BCBL {}\ \BBA {} Creager, K.%
\end{APACrefauthors}%
\unskip\
\newblock
\APACrefYearMonthDay{2012}{}{}.
\newblock
{\BBOQ}\APACrefatitle {Tremor asperities in the transition zone control
  evolution of slow earthquakes} {Tremor asperities in the transition zone
  control evolution of slow earthquakes}.{\BBCQ}
\newblock
\APACjournalVolNumPages{Journal of Geophysical Research}{117}{}{B10301}.
\PrintBackRefs{\CurrentBib}

\bibitem [\protect \citeauthoryear {%
Ghosh%
\ \protect \BOthers {.}}{%
Ghosh%
\ \protect \BOthers {.}}{%
{\protect \APACyear {2010a}}%
}]{%
GHO_2010_GRL}
\APACinsertmetastar {%
GHO_2010_GRL}%
\begin{APACrefauthors}%
Ghosh, A.%
, Vidale, J.%
, Sweet, J.%
, Creager, K.%
, Wech, A.%
\BCBL {}\ \BBA {} Houston, H.%
\end{APACrefauthors}%
\unskip\
\newblock
\APACrefYearMonthDay{2010a}{}{}.
\newblock
{\BBOQ}\APACrefatitle {Tremor bands sweep {Cascadia}} {Tremor bands sweep
  {Cascadia}}.{\BBCQ}
\newblock
\APACjournalVolNumPages{Geophysical Research Letters}{37}{}{L08301}.
\PrintBackRefs{\CurrentBib}

\bibitem [\protect \citeauthoryear {%
Ghosh%
\ \protect \BOthers {.}}{%
Ghosh%
\ \protect \BOthers {.}}{%
{\protect \APACyear {2010b}}%
}]{%
GHO_2010_G3}
\APACinsertmetastar {%
GHO_2010_G3}%
\begin{APACrefauthors}%
Ghosh, A.%
, Vidale, J.%
, Sweet, J.%
, Creager, K.%
, Wech, A.%
, Houston, H.%
\BCBL {}\ \BBA {} Brodsky, E.%
\end{APACrefauthors}%
\unskip\
\newblock
\APACrefYearMonthDay{2010b}{}{}.
\newblock
{\BBOQ}\APACrefatitle {Rapid, continuous streaking of tremor in {Cascadia}}
  {Rapid, continuous streaking of tremor in {Cascadia}}.{\BBCQ}
\newblock
\APACjournalVolNumPages{Geochemistry, Geophysics, Geosystems}{11}{}{Q12010}.
\PrintBackRefs{\CurrentBib}

\bibitem [\protect \citeauthoryear {%
Hyndman%
, McCrory%
, Wech%
, Kao%
\BCBL {}\ \BBA {} J.%
}{%
Hyndman%
\ \protect \BOthers {.}}{%
{\protect \APACyear {2015}}%
}]{%
HYN_2015}
\APACinsertmetastar {%
HYN_2015}%
\begin{APACrefauthors}%
Hyndman, R.%
, McCrory, P.%
, Wech, A.%
, Kao, H.%
\BCBL {}\ \BBA {} J., A.%
\end{APACrefauthors}%
\unskip\
\newblock
\APACrefYearMonthDay{2015}{}{}.
\newblock
{\BBOQ}\APACrefatitle {{Cascadia} subducting plate fluids channelled to
  fore-arc mantle corner: {ETS} and silica deposition} {{Cascadia} subducting
  plate fluids channelled to fore-arc mantle corner: {ETS} and silica
  deposition}.{\BBCQ}
\newblock
\APACjournalVolNumPages{Journal of Geophysical Research Solid
  Earth}{120}{}{4344-4358}.
\PrintBackRefs{\CurrentBib}

\bibitem [\protect \citeauthoryear {%
Ide%
}{%
Ide%
}{%
{\protect \APACyear {2012}}%
}]{%
IDE_2012}
\APACinsertmetastar {%
IDE_2012}%
\begin{APACrefauthors}%
Ide, S.%
\end{APACrefauthors}%
\unskip\
\newblock
\APACrefYearMonthDay{2012}{}{}.
\newblock
{\BBOQ}\APACrefatitle {Variety and spatial heterogeneity of tectonic tremor
  worldwide} {Variety and spatial heterogeneity of tectonic tremor
  worldwide}.{\BBCQ}
\newblock
\APACjournalVolNumPages{Journal of Geophysical Research}{117}{}{B03302}.
\PrintBackRefs{\CurrentBib}

\bibitem [\protect \citeauthoryear {%
Ide%
, Shelly%
\BCBL {}\ \BBA {} Beroza%
}{%
Ide%
\ \protect \BOthers {.}}{%
{\protect \APACyear {2007}}%
}]{%
IDE_2007_GRL}
\APACinsertmetastar {%
IDE_2007_GRL}%
\begin{APACrefauthors}%
Ide, S.%
, Shelly, D.%
\BCBL {}\ \BBA {} Beroza, G.%
\end{APACrefauthors}%
\unskip\
\newblock
\APACrefYearMonthDay{2007}{}{}.
\newblock
{\BBOQ}\APACrefatitle {Mechanism of deep low frequency earthquakes: Further
  evidence that deep non-volcanic tremor is generated by shear slip on the
  plate interface} {Mechanism of deep low frequency earthquakes: Further
  evidence that deep non-volcanic tremor is generated by shear slip on the
  plate interface}.{\BBCQ}
\newblock
\APACjournalVolNumPages{Geophysical Research Letters}{34}{}{L03308}.
\PrintBackRefs{\CurrentBib}

\bibitem [\protect \citeauthoryear {%
Kao%
, Shan%
, Dragert%
\BCBL {}\ \BBA {} Rogers%
}{%
Kao%
\ \protect \BOthers {.}}{%
{\protect \APACyear {2009}}%
}]{%
KAO_2009}
\APACinsertmetastar {%
KAO_2009}%
\begin{APACrefauthors}%
Kao, H.%
, Shan, S\BHBI J.%
, Dragert, H.%
\BCBL {}\ \BBA {} Rogers, G.%
\end{APACrefauthors}%
\unskip\
\newblock
\APACrefYearMonthDay{2009}{}{}.
\newblock
{\BBOQ}\APACrefatitle {Northern {Cascadia} episodic tremor and slip: A decade
  of tremor observations from 1997 to 2007} {Northern {Cascadia} episodic
  tremor and slip: A decade of tremor observations from 1997 to 2007}.{\BBCQ}
\newblock
\APACjournalVolNumPages{Journal of Geophysical Research}{114}{}{B00A12}.
\PrintBackRefs{\CurrentBib}

\bibitem [\protect \citeauthoryear {%
Kao%
\ \protect \BOthers {.}}{%
Kao%
\ \protect \BOthers {.}}{%
{\protect \APACyear {2006}}%
}]{%
KAO_2006}
\APACinsertmetastar {%
KAO_2006}%
\begin{APACrefauthors}%
Kao, H.%
, Shan, S\BHBI J.%
, Dragert, H.%
, Rogers, G.%
, Cassidy, J.%
, Wang, K.%
\BDBL {}Ramachandran, K.%
\end{APACrefauthors}%
\unskip\
\newblock
\APACrefYearMonthDay{2006}{}{}.
\newblock
{\BBOQ}\APACrefatitle {Spatial-temporal patterns of seismic tremors in northern
  {Cascadia}} {Spatial-temporal patterns of seismic tremors in northern
  {Cascadia}}.{\BBCQ}
\newblock
\APACjournalVolNumPages{Journal of Geophysical Research}{111}{}{B03309}.
\PrintBackRefs{\CurrentBib}

\bibitem [\protect \citeauthoryear {%
La~Rocca%
\ \protect \BOthers {.}}{%
La~Rocca%
\ \protect \BOthers {.}}{%
{\protect \APACyear {2009}}%
}]{%
LAR_2009}
\APACinsertmetastar {%
LAR_2009}%
\begin{APACrefauthors}%
La~Rocca, M.%
, Creager, K.%
, Galluzzo, D.%
, Malone, S.%
, Vidale, J.%
, Sweet, J.%
\BCBL {}\ \BBA {} Wech, A.%
\end{APACrefauthors}%
\unskip\
\newblock
\APACrefYearMonthDay{2009}{}{}.
\newblock
{\BBOQ}\APACrefatitle {{Cascadia} tremor located near plate interface
  constrained by {S} minus {P} wave times} {{Cascadia} tremor located near
  plate interface constrained by {S} minus {P} wave times}.{\BBCQ}
\newblock
\APACjournalVolNumPages{Science}{323}{}{620-623}.
\PrintBackRefs{\CurrentBib}

\bibitem [\protect \citeauthoryear {%
McCrory%
, Blair%
, Oppenheimer%
\BCBL {}\ \BBA {} Walter%
}{%
McCrory%
\ \protect \BOthers {.}}{%
{\protect \APACyear {2006}}%
}]{%
MCC_2006}
\APACinsertmetastar {%
MCC_2006}%
\begin{APACrefauthors}%
McCrory, P.%
, Blair, J.%
, Oppenheimer, D.%
\BCBL {}\ \BBA {} Walter, S.%
\end{APACrefauthors}%
\unskip\
\newblock
\APACrefYearMonthDay{2006}{}{}.
\newblock
\APACrefbtitle {Depth to the {Juan} de {Fuca} slab beneath the {Cascadia}
  subduction margin - A {3-D} model sorting earthquakes} {Depth to the {Juan}
  de {Fuca} slab beneath the {Cascadia} subduction margin - a {3-D} model
  sorting earthquakes}\ \APACbVolEdTR{}{\BTR{}\ \BNUM\ Series 91}.
\newblock
\APACaddressInstitution{}{U.S. Geological Survey}.
\PrintBackRefs{\CurrentBib}

\bibitem [\protect \citeauthoryear {%
Merrill%
, Bostock%
, Peacock%
, Calvert%
\BCBL {}\ \BBA {} Christensen%
}{%
Merrill%
\ \protect \BOthers {.}}{%
{\protect \APACyear {2020}}%
}]{%
MER_2020}
\APACinsertmetastar {%
MER_2020}%
\begin{APACrefauthors}%
Merrill, R.%
, Bostock, M\BPBI G.%
, Peacock, S.%
, Calvert, A\BPBI J.%
\BCBL {}\ \BBA {} Christensen, N\BPBI I.%
\end{APACrefauthors}%
\unskip\
\newblock
\APACrefYearMonthDay{2020}{}{}.
\newblock
{\BBOQ}\APACrefatitle {A double difference tomography study of the {Washington}
  {Forearc}: Does {Siletzia} control crustal seismicity?} {A double difference
  tomography study of the {Washington} {Forearc}: Does {Siletzia} control
  crustal seismicity?}{\BBCQ}
\newblock
\APACjournalVolNumPages{Journal of Geophysical Research: Solid
  Earth}{125}{}{e2020JB019750}.
\PrintBackRefs{\CurrentBib}

\bibitem [\protect \citeauthoryear {%
Nakata%
, Suda%
\BCBL {}\ \BBA {} Tsuruoka%
}{%
Nakata%
\ \protect \BOthers {.}}{%
{\protect \APACyear {2008}}%
}]{%
NAK_2008}
\APACinsertmetastar {%
NAK_2008}%
\begin{APACrefauthors}%
Nakata, R.%
, Suda, N.%
\BCBL {}\ \BBA {} Tsuruoka, H.%
\end{APACrefauthors}%
\unskip\
\newblock
\APACrefYearMonthDay{2008}{}{}.
\newblock
{\BBOQ}\APACrefatitle {Non-volcanic tremor resulting from the combined effect
  of {Earth} tides and slow slip events} {Non-volcanic tremor resulting from
  the combined effect of {Earth} tides and slow slip events}.{\BBCQ}
\newblock
\APACjournalVolNumPages{Nature Geoscience}{1}{}{676-678}.
\PrintBackRefs{\CurrentBib}

\bibitem [\protect \citeauthoryear {%
Nikulin%
, Levin%
\BCBL {}\ \BBA {} Park%
}{%
Nikulin%
\ \protect \BOthers {.}}{%
{\protect \APACyear {2009}}%
}]{%
NIK_2009}
\APACinsertmetastar {%
NIK_2009}%
\begin{APACrefauthors}%
Nikulin, A.%
, Levin, V.%
\BCBL {}\ \BBA {} Park, J.%
\end{APACrefauthors}%
\unskip\
\newblock
\APACrefYearMonthDay{2009}{}{}.
\newblock
{\BBOQ}\APACrefatitle {Receiver function study of the {Cascadia} megathrust:
  Evidence for localized serpentinization} {Receiver function study of the
  {Cascadia} megathrust: Evidence for localized serpentinization}.{\BBCQ}
\newblock
\APACjournalVolNumPages{Geochemistry, Geophysics, Geosystems}{10}{}{Q07004}.
\PrintBackRefs{\CurrentBib}

\bibitem [\protect \citeauthoryear {%
Obara%
}{%
Obara%
}{%
{\protect \APACyear {2002}}%
}]{%
OBA_2002}
\APACinsertmetastar {%
OBA_2002}%
\begin{APACrefauthors}%
Obara, K.%
\end{APACrefauthors}%
\unskip\
\newblock
\APACrefYearMonthDay{2002}{}{}.
\newblock
{\BBOQ}\APACrefatitle {Nonvolcanic deep tremor associated with subduction in
  southwest {Japan}} {Nonvolcanic deep tremor associated with subduction in
  southwest {Japan}}.{\BBCQ}
\newblock
\APACjournalVolNumPages{Science}{296}{5573}{1679-1681}.
\PrintBackRefs{\CurrentBib}

\bibitem [\protect \citeauthoryear {%
Peacock%
}{%
Peacock%
}{%
{\protect \APACyear {2009}}%
}]{%
PEA_2009}
\APACinsertmetastar {%
PEA_2009}%
\begin{APACrefauthors}%
Peacock, S.%
\end{APACrefauthors}%
\unskip\
\newblock
\APACrefYearMonthDay{2009}{}{}.
\newblock
{\BBOQ}\APACrefatitle {Thermal and metamorphic environment of subduction zone
  episodic tremor and slip} {Thermal and metamorphic environment of subduction
  zone episodic tremor and slip}.{\BBCQ}
\newblock
\APACjournalVolNumPages{Journal of Geophysical Research}{114}{}{B00A07}.
\PrintBackRefs{\CurrentBib}

\bibitem [\protect \citeauthoryear {%
Preston%
, Creager%
, Crosson%
, Brocher%
\BCBL {}\ \BBA {} Trehu%
}{%
Preston%
\ \protect \BOthers {.}}{%
{\protect \APACyear {2003}}%
}]{%
PRE_2003}
\APACinsertmetastar {%
PRE_2003}%
\begin{APACrefauthors}%
Preston, L.%
, Creager, K.%
, Crosson, R.%
, Brocher, T.%
\BCBL {}\ \BBA {} Trehu, A.%
\end{APACrefauthors}%
\unskip\
\newblock
\APACrefYearMonthDay{2003}{}{}.
\newblock
{\BBOQ}\APACrefatitle {Intraslab earthquakes: Dehydration of the {Cascadia
  slab}} {Intraslab earthquakes: Dehydration of the {Cascadia slab}}.{\BBCQ}
\newblock
\APACjournalVolNumPages{Science}{302}{}{1197-1200}.
\PrintBackRefs{\CurrentBib}

\bibitem [\protect \citeauthoryear {%
Rogers%
\ \BBA {} Dragert%
}{%
Rogers%
\ \BBA {} Dragert%
}{%
{\protect \APACyear {2003}}%
}]{%
ROG_2003}
\APACinsertmetastar {%
ROG_2003}%
\begin{APACrefauthors}%
Rogers, G.%
\BCBT {}\ \BBA {} Dragert, H.%
\end{APACrefauthors}%
\unskip\
\newblock
\APACrefYearMonthDay{2003}{}{}.
\newblock
{\BBOQ}\APACrefatitle {Tremor and slip on the {Cascadia} subduction zone: The
  chatter of silent slip} {Tremor and slip on the {Cascadia} subduction zone:
  The chatter of silent slip}.{\BBCQ}
\newblock
\APACjournalVolNumPages{Science}{300}{5627}{1942-1943}.
\PrintBackRefs{\CurrentBib}

\bibitem [\protect \citeauthoryear {%
Rousseeuw%
\ \BBA {} Croux%
}{%
Rousseeuw%
\ \BBA {} Croux%
}{%
{\protect \APACyear {1993}}%
}]{%
ROU_1993}
\APACinsertmetastar {%
ROU_1993}%
\begin{APACrefauthors}%
Rousseeuw, P.%
\BCBT {}\ \BBA {} Croux, C.%
\end{APACrefauthors}%
\unskip\
\newblock
\APACrefYearMonthDay{1993}{}{}.
\newblock
{\BBOQ}\APACrefatitle {Alternatives to the median absolute deviation}
  {Alternatives to the median absolute deviation}.{\BBCQ}
\newblock
\APACjournalVolNumPages{Journal of the American Statistical
  Association}{88}{424}{1273-1283}.
\PrintBackRefs{\CurrentBib}

\bibitem [\protect \citeauthoryear {%
Rowe%
\ \BBA {} Moore%
}{%
Rowe%
\ \BBA {} Moore%
}{%
{\protect \APACyear {2013}}%
}]{%
ROW_2013}
\APACinsertmetastar {%
ROW_2013}%
\begin{APACrefauthors}%
Rowe, C.%
\BCBT {}\ \BBA {} Moore, F., J.~Remitti.%
\end{APACrefauthors}%
\unskip\
\newblock
\APACrefYearMonthDay{2013}{}{}.
\newblock
{\BBOQ}\APACrefatitle {The thickness of subduction plate boundary faults from
  the seafloor into the seismogenic zone} {The thickness of subduction plate
  boundary faults from the seafloor into the seismogenic zone}.{\BBCQ}
\newblock
\APACjournalVolNumPages{Geology}{41}{9}{991-994}.
\PrintBackRefs{\CurrentBib}

\bibitem [\protect \citeauthoryear {%
Royer%
\ \BBA {} Bostock%
}{%
Royer%
\ \BBA {} Bostock%
}{%
{\protect \APACyear {2014}}%
}]{%
ROY_2014}
\APACinsertmetastar {%
ROY_2014}%
\begin{APACrefauthors}%
Royer, A.%
\BCBT {}\ \BBA {} Bostock, M.%
\end{APACrefauthors}%
\unskip\
\newblock
\APACrefYearMonthDay{2014}{}{}.
\newblock
{\BBOQ}\APACrefatitle {A comparative study of low frequency earthquake
  templates in northern {Cascadia}} {A comparative study of low frequency
  earthquake templates in northern {Cascadia}}.{\BBCQ}
\newblock
\APACjournalVolNumPages{Earth and Planetary Science Letters}{402}{}{247-256}.
\PrintBackRefs{\CurrentBib}

\bibitem [\protect \citeauthoryear {%
Rubin%
\ \BBA {} Armbruster%
}{%
Rubin%
\ \BBA {} Armbruster%
}{%
{\protect \APACyear {2013}}%
}]{%
RUB_2013}
\APACinsertmetastar {%
RUB_2013}%
\begin{APACrefauthors}%
Rubin, A.%
\BCBT {}\ \BBA {} Armbruster, J.%
\end{APACrefauthors}%
\unskip\
\newblock
\APACrefYearMonthDay{2013}{}{}.
\newblock
{\BBOQ}\APACrefatitle {Imaging slow slip fronts in {Cascadia} with high
  precision cross-station tremor locations} {Imaging slow slip fronts in
  {Cascadia} with high precision cross-station tremor locations}.{\BBCQ}
\newblock
\APACjournalVolNumPages{Geochemistry Geophysics Geosystems}{14}{}{5371-5392}.
\PrintBackRefs{\CurrentBib}

\bibitem [\protect \citeauthoryear {%
Sammis%
\ \BBA {} Bostock%
}{%
Sammis%
\ \BBA {} Bostock%
}{%
{\protect \APACyear {2021}}%
}]{%
SAM_2021}
\APACinsertmetastar {%
SAM_2021}%
\begin{APACrefauthors}%
Sammis, C.%
\BCBT {}\ \BBA {} Bostock, M.%
\end{APACrefauthors}%
\unskip\
\newblock
\APACrefYearMonthDay{2021}{}{}.
\newblock
{\BBOQ}\APACrefatitle {A granular jamming model for low-frequency earthquakes}
  {A granular jamming model for low-frequency earthquakes}.{\BBCQ}
\newblock
\APACjournalVolNumPages{Journal of Geophysical Research: Solid
  Earth}{126}{}{e2021JB021963}.
\PrintBackRefs{\CurrentBib}

\bibitem [\protect \citeauthoryear {%
Schimmel%
\ \BBA {} Paulssen%
}{%
Schimmel%
\ \BBA {} Paulssen%
}{%
{\protect \APACyear {1997}}%
}]{%
SCH_1997}
\APACinsertmetastar {%
SCH_1997}%
\begin{APACrefauthors}%
Schimmel, M.%
\BCBT {}\ \BBA {} Paulssen, H.%
\end{APACrefauthors}%
\unskip\
\newblock
\APACrefYearMonthDay{1997}{}{}.
\newblock
{\BBOQ}\APACrefatitle {Noise reduction and detection of weak, coherent signals
  through phase-weighted stacks} {Noise reduction and detection of weak,
  coherent signals through phase-weighted stacks}.{\BBCQ}
\newblock
\APACjournalVolNumPages{Geophysical Journal International}{130}{}{487-505}.
\PrintBackRefs{\CurrentBib}

\bibitem [\protect \citeauthoryear {%
Shelly%
, Beroza%
\BCBL {}\ \BBA {} Ide%
}{%
Shelly%
\ \protect \BOthers {.}}{%
{\protect \APACyear {2007}}%
}]{%
SHE_2007_nature}
\APACinsertmetastar {%
SHE_2007_nature}%
\begin{APACrefauthors}%
Shelly, D.%
, Beroza, G.%
\BCBL {}\ \BBA {} Ide, S.%
\end{APACrefauthors}%
\unskip\
\newblock
\APACrefYearMonthDay{2007}{}{}.
\newblock
{\BBOQ}\APACrefatitle {Non-volcanic tremor and low-frequency earthquake swarms}
  {Non-volcanic tremor and low-frequency earthquake swarms}.{\BBCQ}
\newblock
\APACjournalVolNumPages{Nature}{446}{}{305-307}.
\PrintBackRefs{\CurrentBib}

\bibitem [\protect \citeauthoryear {%
Shelly%
, Beroza%
, Ide%
\BCBL {}\ \BBA {} Nakamula%
}{%
Shelly%
\ \protect \BOthers {.}}{%
{\protect \APACyear {2006}}%
}]{%
SHE_2006}
\APACinsertmetastar {%
SHE_2006}%
\begin{APACrefauthors}%
Shelly, D.%
, Beroza, G.%
, Ide, S.%
\BCBL {}\ \BBA {} Nakamula, S.%
\end{APACrefauthors}%
\unskip\
\newblock
\APACrefYearMonthDay{2006}{}{}.
\newblock
{\BBOQ}\APACrefatitle {Low-frequency earthquakes in {Shikoku}, {Japan}, and
  their relationship to episodic tremor and slip} {Low-frequency earthquakes in
  {Shikoku}, {Japan}, and their relationship to episodic tremor and
  slip}.{\BBCQ}
\newblock
\APACjournalVolNumPages{Nature}{442}{}{188-192}.
\PrintBackRefs{\CurrentBib}

\bibitem [\protect \citeauthoryear {%
Sweet%
, Creager%
, Houston%
\BCBL {}\ \BBA {} Chestler%
}{%
Sweet%
\ \protect \BOthers {.}}{%
{\protect \APACyear {2019}}%
}]{%
SWE_2019}
\APACinsertmetastar {%
SWE_2019}%
\begin{APACrefauthors}%
Sweet, J.%
, Creager, K.%
, Houston, H.%
\BCBL {}\ \BBA {} Chestler, S.%
\end{APACrefauthors}%
\unskip\
\newblock
\APACrefYearMonthDay{2019}{}{}.
\newblock
{\BBOQ}\APACrefatitle {Variations in {Cascadia} low-frequency earthquake
  behavior with downdip distance} {Variations in {Cascadia} low-frequency
  earthquake behavior with downdip distance}.{\BBCQ}
\newblock
\APACjournalVolNumPages{Geochemistry, Geophysics, Geosystems}{20}{}{1202-1217}.
\PrintBackRefs{\CurrentBib}

\bibitem [\protect \citeauthoryear {%
Thomas%
, Nadeau%
\BCBL {}\ \BBA {} B\"urgmann%
}{%
Thomas%
\ \protect \BOthers {.}}{%
{\protect \APACyear {2009}}%
}]{%
THO_2009}
\APACinsertmetastar {%
THO_2009}%
\begin{APACrefauthors}%
Thomas, A.%
, Nadeau, R.%
\BCBL {}\ \BBA {} B\"urgmann, R.%
\end{APACrefauthors}%
\unskip\
\newblock
\APACrefYearMonthDay{2009}{}{}.
\newblock
{\BBOQ}\APACrefatitle {Tremor-tide correlations and near-lithostatic pore
  pressure on the deep {San} {Andreas} fault} {Tremor-tide correlations and
  near-lithostatic pore pressure on the deep {San} {Andreas} fault}.{\BBCQ}
\newblock
\APACjournalVolNumPages{Nature}{462}{}{1048-1051}.
\PrintBackRefs{\CurrentBib}

\bibitem [\protect \citeauthoryear {%
Wech%
\ \BBA {} Creager%
}{%
Wech%
\ \BBA {} Creager%
}{%
{\protect \APACyear {2007}}%
}]{%
WEC_2007}
\APACinsertmetastar {%
WEC_2007}%
\begin{APACrefauthors}%
Wech, A.%
\BCBT {}\ \BBA {} Creager, K.%
\end{APACrefauthors}%
\unskip\
\newblock
\APACrefYearMonthDay{2007}{}{}.
\newblock
{\BBOQ}\APACrefatitle {Cascadia tremor polarization evidence for plate
  interface slip} {Cascadia tremor polarization evidence for plate interface
  slip}.{\BBCQ}
\newblock
\APACjournalVolNumPages{Geophysical Research Letters}{34}{}{L22306}.
\PrintBackRefs{\CurrentBib}

\bibitem [\protect \citeauthoryear {%
Wech%
\ \BBA {} Creager%
}{%
Wech%
\ \BBA {} Creager%
}{%
{\protect \APACyear {2008}}%
}]{%
WEC_2008}
\APACinsertmetastar {%
WEC_2008}%
\begin{APACrefauthors}%
Wech, A.%
\BCBT {}\ \BBA {} Creager, K.%
\end{APACrefauthors}%
\unskip\
\newblock
\APACrefYearMonthDay{2008}{}{}.
\newblock
{\BBOQ}\APACrefatitle {Automated detection and location of {Cascadia} tremor}
  {Automated detection and location of {Cascadia} tremor}.{\BBCQ}
\newblock
\APACjournalVolNumPages{Geophysical Research Letters}{35}{}{L20302}.
\PrintBackRefs{\CurrentBib}

\bibitem [\protect \citeauthoryear {%
Wessel%
\ \BBA {} Smith%
}{%
Wessel%
\ \BBA {} Smith%
}{%
{\protect \APACyear {1991}}%
}]{%
WES_1991}
\APACinsertmetastar {%
WES_1991}%
\begin{APACrefauthors}%
Wessel, P.%
\BCBT {}\ \BBA {} Smith, W\BPBI H\BPBI F.%
\end{APACrefauthors}%
\unskip\
\newblock
\APACrefYearMonthDay{1991}{}{}.
\newblock
{\BBOQ}\APACrefatitle {Free software helps map and display data} {Free software
  helps map and display data}.{\BBCQ}
\newblock
\APACjournalVolNumPages{EOS Trans. AGU}{72}{}{441}.
\PrintBackRefs{\CurrentBib}

\end{thebibliography}

%
%
%
%
%

\end{document}